\documentclass[amsthm]{elsart}
\usepackage{amssymb,url}

\usepackage{yjsco}
\usepackage{natbib}
\usepackage[dvipdfmx]{graphics}
\usepackage{fancybox,amsmath,amsbsy,ascmac,graphicx} 

\usepackage{hyperref}

\DeclareMathOperator{\C}{\mathbb{C}}
\DeclareMathOperator{\N}{\mathbb{N}}
\DeclareMathOperator{\V}{\mathbb{V}}
\DeclareMathOperator{\A}{\mathbb{A}}
\DeclareMathOperator{\Q}{\mathbb{Q}}
\DeclareMathOperator{\B}{\mathbb{B}}

\newlength{\myheight}
\newlength{\myheighta}
\DeclareMathOperator{\LL}{LL}
\DeclareMathOperator{\TList}{TList}
\DeclareMathOperator{\GList}{GList}
\DeclareMathOperator{\MList}{MList}
\DeclareMathOperator{\SList}{SList}
\DeclareMathOperator{\LList}{LList}
\DeclareMathOperator{\TT}{TT}
\DeclareMathOperator{\FL}{FL}
\DeclareMathOperator{\CL}{CL}
\DeclareMathOperator{\RR}{RR}
\DeclareMathOperator{\EL}{EL}
\DeclareMathOperator{\UU}{UU}

\DeclareMathOperator{\CT}{CT}
\DeclareMathOperator{\het}{ht}
\DeclareMathOperator{\mdeg}{mdeg}

\DeclareMathOperator{\hc}{hc}
\DeclareMathOperator{\hm}{hm}

\DeclareMathOperator{\Mono}{Mono}
\DeclareMathOperator{\CV}{{\cal CV}}

\newcommand{\Set}[1]{\left\{#1\right\}}
\newcommand{\setDef}[2]{{#1}\left|\,\vphantom{#1}{#2}\right.}
\newcommand{\SetDef}[2]{\Set{\setDef{#1}{#2}}}

\def\SEN{\hrule width \textwidth}

\newtheorem{notation}[thm]{Notation}

\begin{document}

\begin{frontmatter}

\title{Algebraic Local Cohomology with Parameters and Parametric Standard Bases for Zero-Dimensional Ideals}

\thanks{A part of this work has been supported by Grant-in-Aid for Scientific ResearchiCj (No. 2454016201).}
 
\author{Katsusuke Nabeshima}
\address{Institute of Socio-Arts and Sciences, Tokushima University, \\1-1, Minamijosanjima, Tokushima, JAPAN}
\ead{nabeshima@tokushima-u.ac.jp}

\author{Shinichi Tajima}
\address{Graduate School of Pure and Applied Sciences, University of Tsukuba, \\1-1-1, Tennoudai, Tsukuba, JAPAN}
\ead{tajima@math.tsukuba.ac.jp}



\begin{abstract}
 A computation method of algebraic local cohomology with parameters, 
associated with zero-dimensional ideal with parameter, is
 introduced. This computation method gives us in particular a decomposition of the
 parameter space depending on the structure of algebraic local cohomology
classes. This decomposition informs us several  properties of input ideals and 
the output of our algorithm completely describes the multiplicity structure of 
input ideals. 
 An efficient algorithm for
 computing a parametric standard basis of a given zero-dimensional ideal,
 with respect to an arbitrary local term order, is also described as an
application of the computation method. The algorithm can always output
 ``reduced'' standard basis of a given zero-dimensional ideal, even if the
 zero-dimensional ideal has parameters.
\end{abstract}


\end{frontmatter}


\section{Introduction}
Local cohomology was introduced by A. Grothendieck in \citep{Grot}. Subsequent development to a great extent has been motivated by Grothendieck's ideas \cite{BS98,Lyu02}. Nowadays, local cohomology is a key ingredient in algebraic geometry, commutative algebra, topology and D-modules, and is a fundamental tool for applications in several fields.

In \citep{TNN09}, we  proposed, with Y. Nakamura, an algorithmic method to compute algebraic local cohomology classes, supported at a point, associated with a given zero-dimensional ideal. We described therein an efficient method for computing standard bases of zero-dimensional ideals, that utilize algebraic local cohomology classes. The underlying idea of the proposed method comes from the fact
that algebraic local cohomology classes can completely describe the multiplicity structure of a zero-dimensional ideal via the Grothendieck local duality theorem. 
More recently in our result of ISSAC2014 \citep{NT14}, we considered the Jacobi ideal, with deformation parameter, of a semi-quasihomogeneous hypersurface isolated singularity. 
By adopting the same approach presented in \cite{TNN09}, we constructed an algorithm for computing algebraic local cohomology classes, with parameters, that are annihilated by the Jacobi ideal. As an application, we obtained a new method to compute parametric standard bases of Jacobi ideals  associated with a deformation of semi-quasihomogeneous hypersurface isolated singularities.

In this paper, we address the problem of finding  an effective method to treat algebraic local cohomology classes with parameters associated with a  given zero-dimensional ideal with parameters, that works in general cases.

In order to state precisely the problem, let $X$ be an open neighborhood of the origin $O$ of the $n$-dimensional complex space $\mathbb{C}^n$ with coordinates $x=(x_1,x_2,\ldots,x_n)$. We assume that a set $F$ of $p$ polynomials $f_1,f_2,\ldots,f_p$ in $(\mathbb{C}[t_1,\ldots,t_m])[x]$ satisfying generically $\{a\in X|f_1(a)=\cdots=f_p(a)=0\}=\{O\}$ are given where $t_1,\ldots, t_m$ are parameters.  Let $H_F$ be a set of algebraic local cohomology classes supported at the origin that are annihilated by the ideal generated by $F$. Then $H_F$ is a finite-dimensional vector space if and only if the ideal $\langle F \rangle$ generated by $F$ is zero-dimensional in the rings of formal power series. 
In such cases, there is a possibility that $\{a\in X|f_1(a)=\cdots=f_p(a)=0\}\neq \{O\}$ (the same meaning is that $\langle F \rangle$ is not zero-dimensional)  for some values of parameters, because of parameters. As our aim is to construct algorithms for studying the structure of $H_F$ and the multiplicity structure of $\langle F \rangle$ on $X$, it is necessary, beforehand if possible, to detect these values of parameters, that constitute constructible sets, from the parameter space for computing algebraic local cohomology classes.

In the first part of this paper, we introduce a new notion of parametric local cohomology system as an analogue of comprehensive system to deal with parametric problems. We describe a new effective method to compute parametric local cohomology systems. The resulting algorithms compute in particular a suitable decomposition of parameter space to a finite set of constructible sets according to the structure of algebraic local cohomology classes in question. 
The key of the algorithm for decomposing is the use of a comprehensive Gr\"obner bases computation in a polynomial ring with parameters.
The algorithms for computing bases of $H_F$, is designed as dynamic algorithm in consideration of computational efficiency. 
The output of our algorithm, has the abundant information of the input ideal and provides a complete description of the multiplicity structures of parametric zero-dimensional ideals.

In the second part of this paper, we describe algorithms for computing parametric standard bases as an application of parametric local cohomology systems. We show that the use of algebraic local cohomology provides an efficient algorithm for computing standard bases. 
Furthermore, the use of algebraic local cohomology transforms a standard basis of a dimensional ideal $\langle F \rangle$ with respect to any given local term order into a standard basis with respect to any other ordering, without computing the standard basis, again.
In general, the computation complexity of standard bases, is strongly influenced by the term order, like Gr\"obner bases computation. Thus, this property is useful to compute a standard basis.

Especially, our algorithm can output always ``reduced'' standard basis of a given zero-dimensional ideal, even if $F$ has parameters. Note that, an algorithm implemented in the computer algebra system \verb|Singular| \citep{DGPS} that compute standard bases  does not enjoy this property. Moreover, in general, comprehensive Gr\"obner basis \citep{Na12,wv92} in a polynomial ring does not have this property, too.

As we mentioned above, there are several applications of algebraic local cohomology. For examples, our  algorithm can be used to analyze properties of singularities and deformations of Artin algebra \citep{Iar78, Iar84}. It is a powerful tool to study several problems relevant to zero-dimensional ideals.

All algorithms in this paper, have been implemented in the computer algebra system \verb|Risa/Asir| \citep{NT92}. 

This paper is organized as follows. Section 2 briefly reviews algebraic local cohomology, and gives notations and definitions used in this paper. Section 3 is the discussion of the new algorithm for algebraic local cohomology classes with parameters. This section is the main part of this paper. Section 4 gives  algorithms for computing parametric standard bases for a given zero-dimensional ideals.
\section{Preliminaries}
In this section, first we briefly review algebraic local cohomology. Second, we introduce a term order for computing algebraic local cohomology classes and  algebraically constructible sets, which will be exploited several times in this paper. Throughout this paper, we use the notation $x$ as the abbreviation of $n$ variables $x_1,\ldots, x_n$.  The set of natural number $\N$ includes zero. 
$K$ is the field of rational numbers $\Q$ or the field of complex numbers $\C$. 
\subsection{Algebraic local cohomology}

Let $H^n_{[O]}(K[x])$ denote the set of algebraic local cohomology classes supported at the origin $O$ with coefficients in $K$, defined by 
$$\displaystyle H^n_{[O]}(K[x]):=\lim_{k\rightarrow \infty}Ext^n_{K[x]}(K[x]/\langle x_1,x_2,..,x_n\rangle^k,K[x])$$
 where $\langle x_1,x_2,$ $\ldots,x_n \rangle$ is the maximal ideal generated by $x_1,x_2,\ldots,x_n$.

Let $X$ be a neighborhood of the origin $O$ of $K^n$. Consider the pair $(X,X-O)$ and its relative \v{C}ech covering. Then, any section of $H^n_{[O]}(K[x])$ can be represented as an element of relative \v{C}ech cohomology. 
We use the notation $\sum c_{\lambda} \left[\frac{1}{x^{\lambda+1}}\right]$  for representing an algebraic local cohomology class in $H^n_{[O]}(K[x])$ where $c_{\lambda}\in K$, $x^{\lambda+1}=x_1^{\lambda_1+1}x_2^{\lambda_2+1}\cdots $ $x_n^{\lambda_n+1}$ with $\lambda=(\lambda_1,\lambda_2,\ldots,\lambda_n)\in \N^n$. Note that the multiplication is defined as 
$$ \displaystyle x^{\alpha}\left[\frac{1}{x^{\lambda+1}}\right]
:=\left\{
  \begin{array}{ll}
\displaystyle \left[\frac{1}{x^{\lambda+1-\alpha}} \right], \ \ \ \ \ \ \ \lambda_i\ge \alpha_i, i=1,\ldots,n,\\
\ \\
 0,  \ \ \ \ \ \ \ \  \ \ \ \ \ \ \ \ \text{otherwise}, 
\end{array}
\right.
$$
where $\alpha=(\alpha_1,\ldots,\alpha_n) \in \N^n$ and $\lambda+1-\alpha=(\lambda_1+1-\alpha_1,\ldots, \lambda_n+1-\alpha_n)$. 

We represent an algebraic local cohomology class $\sum c_{\lambda}\left[\frac{1}{x^{\lambda+1}}\right]$ as a polynomial in $n$ variables $\sum c_{\lambda}\xi^\lambda$ to manipulate algebraic local cohomology classes efficiently (on computer),  where $\xi$ is the abbreviation of $n$ variables $\xi_1,\xi_2,\ldots,\xi_n$. We call this representation ``{\bf polynomial representation}''. For example, let $\psi=\left[\frac{4}{x_1^3x_2^4}\right]+\left[\frac{5}{x_1^2x_2^3}\right]$ be an algebraic local cohomology class where $x_1, x_2$ are variables. Then, the polynomial representation of $\psi$, is $4\xi_1^2\xi_2^3+5\xi_1\xi_2^2$ where variables $\xi_1, \xi_2$ are corresponding to variables $x_1, x_2$. That is, we have the following table for $n$ variables: 
$$\displaystyle 
\begin{tabular}{|ccc|}
\hline
\rule{0cm}{\myheight} {\bf \v{C}ech representation} & & {\bf polynomial representation}\\ \hline
\vspace{-3.0mm} & & \\
$\displaystyle \sum c_{\lambda}\left[\frac{1}{x_1^{\lambda_1+1}x_2^{\lambda_2+1}\cdots x_n^{\lambda_n+1}}\right]$ & $\longleftrightarrow$ & $\displaystyle \sum c_{\lambda}\xi_1^{\lambda_1}\xi_2^{\lambda_2}\cdots \xi_n^{\lambda_n}$ \vspace{-3.0mm} \\ 
 & & \\ \hline
\end{tabular} 
$$
where $c_{\lambda} \in K$. 
The multiplication for polynomial representation is defined as follows:
$$ \displaystyle x^{\alpha}*\xi^\lambda
:=\left\{
\begin{array}{ll}
\xi^{\lambda-\alpha},\ \ \ \ \ \ \ \lambda_i\ge \alpha_i, i=1,\ldots,n,\\
\ \\
 0,  \ \ \ \ \ \ \ \  \ \ \ \ \ \ \ \ \text{otherwise,} 
\end{array}
\right.
$$
where $\alpha=(\alpha_1,\ldots,\alpha_n) \in \N^n, \lambda=(\lambda_1,\ldots,\lambda_n) \in \N^n$, and $\lambda-\alpha=(\lambda_1-\alpha_1,\ldots, \lambda_n-\alpha_n)$. We use `` $\ast$ '' for polynomial representation. 

After here, we adapt polynomial representation to represent an algebraic local cohomology class. 
We use mainly the following term order to compute algebraic local cohomology classes. 

\begin{defn}\label{monoorder}\normalfont 
For two multi-indices $\lambda=(\lambda_1,\lambda_2,\ldots,\lambda_n)$ and $\lambda^{\prime}=(\lambda^{\prime}_1, \lambda^{\prime}_2,\ldots,\lambda^{\prime}_n)$ in $\N^n$, we denote 
$\xi^{\lambda^{\prime}}\prec \xi^{\lambda} \text{ or } \lambda^{\prime} \prec \lambda$ 
if $|\xi^{\lambda^{\prime}}| < |\xi^{\lambda}|$, or if $|\xi^{\lambda^{\prime}}|=|\xi^{\lambda}|$ and there exists $i, j \in \N$ so that $\lambda^{\prime}_i=\lambda_i$ for $i<j$ and $\lambda^{\prime}_j < \lambda_j$ where $|\xi^{\lambda}|=\sum_{i=1}^n\lambda_i$. In general, this term order is called a {\bf total degree lexicographic} term order.
\end{defn}

For a given algebraic local cohomology class $\psi$ of the form, 
$\psi=c_{\lambda}\xi^{\lambda}+\sum_{\lambda' \prec \lambda}c_{\lambda'}\xi^{\lambda'}$, $c_{\lambda}\neq 0,$ 
we call $\xi^{\lambda}$ the {\bf head term} and $\xi^{\lambda'}$, $\lambda'\prec \lambda$ the {\bf lower terms}. We denote the head term of a cohomology class $\psi$ by $\het(\psi)$. 

\subsection{Strata and specialization}
We use the notation $t$ as the abbreviation of $m$ variables $t_1,\ldots,t_m$. (One can also regard $t$ as parameters.) Let $\bar{K}$ be an algebraic closure field of $K$. 
For $g_1,\ldots,g_q \in K[t]$, $\V(g_1,\ldots,g_q) \subseteq \bar{K}^m$ denotes the affine variety of $g_1,\ldots,g_q$, i.e., $\V(g_1,\ldots,g_q):=\{\bar{a}\in \bar{K}^m |$ $ g_1(\bar{a})=\cdots=g_q(\bar{a})=0\}$ and $\V(0):=\bar{K}^m.$

We use an algebraically constructible set that has a form $\V(g_1,\ldots,g_q) \backslash $ $\V(g_1^{\prime},\ldots,g_{q^{\prime}}^{\prime}) \subseteq \bar{K}^m$ where $g_1,\ldots,g_q,$ $g_1^{\prime},\ldots, g_{q^{\prime}}^{\prime} \in K[t]$.  We call the form $\V(g_1,\ldots,g_q) \backslash $ $\V(g_1^{\prime},\ldots,g_{q^{\prime}}^{\prime})$ a {\bf stratum}. (Notation $\A, \A', \A_1,\ldots, \A_l$ are frequently used to represent strata.)

When we treat with systems of parametric equations, then it is necessary to check consistency of their parametric consistents. In several papers \citep{ksd10, Mo02,ss03}, algorithms for checking consistency have been already introduced. Thus, it is possible to decide whether $\V(Q_1)\backslash \V(Q_2)$ is an empty set or not, by these algorithms where $Q_1, Q_2 \subset K[t]$. The details are in the papers. 

We define the localization of $K[t]$ w.r.t. a stratum $\A\subseteq \bar{K}^m$ as follows: $K[t]_{\A} = \{ \frac{c}{b} \mid c, b \in K[t], b(t) \ne 0 \text{ for } t \in \A \}$. Then, for every $\bar{a} \in \mathbb{A}$, we can define the canonical specialization homomorphism $\sigma_{\bar{a}}: K[t]_{\A}[x] \to \bar{K}[x]$ (or $\sigma_{\bar{a}}: K[t]_{\A}[\xi] \to \bar{K}[\xi]$). When we say that $\sigma_{\bar{a}}(h)$ makes sense for $h \in K(t)[x]$, it has to be understood that $h \in K[t]_{\A}[x]$ for some $\A$ with $\bar{a}\in \A_i$. 
We can regard $\sigma_{\bar{a}}$ as substituting $\bar{a}$ into $m$ variables $t$.

\section{Algebraic local cohomology with Parameters}
Let us assume that a set $F$ of $p$ polynomials $f_1,f_2,\ldots,f_p$ in $(K[t])[x]$ satisfying {\bf generically} $\{a\in X|f_1(a)=\cdots=f_p(a)=0\}=\{O\}$ are given where $X$ is a neighborhood of the origin $O$ of $\bar{K}^n$. Here, we regard $t$ as parameters, and $x$, $\xi$ are the main variables. 

We define a set $ H_F = \cup_{\bar{a} \in \bar{K}^m} H_{\sigma_{\bar{a}}(F)}  $  to be the set of algebraic local cohomology classes in $ K[\xi] $ that are annihilated by the ideal generated by $ F $, where 
 $$  H_{\sigma_{\bar{a}}(F)}  = \{ \psi \in \bar{K}[\xi] \mid  \sigma_{\bar{a}}(f_1)\ast \psi = \sigma_{\bar{a}}(f_2)\ast \psi = \cdots =  \sigma_{\bar{a}}(f_p)\ast \psi = 0 \}. $$
The ideal $ \langle F \rangle $ at $ \bar{a} \in \bar{K}^m $ is a zero-dimensional ideal if and only if  $  H_{\sigma_{\bar{a}}(F)} $ is a finite-dimensional vector space. 
In this section we describe an algorithm for computing bases of the vector space $H_F$.  More precisely, we describe algorithms for computing parametric local cohomology systems (see Definition~\ref{55} in this section).

The new algorithm consists of the following three parts.
\begin{enumerate}
\item[(1)]Decompose the parameter space $\bar{K}^m$ into safe strata and danger strata. 
\item[(2)]Compute bases of the vector space $H_F$ on safe strata.
\item[(3)]Compute bases of the vector space $H_F$ on danger strata.
\end{enumerate}

\subsection{An algorithm for testing dimensions of a parametric ideal}
Since polynomials  $f_1,f_2,\ldots,f_p$ have parameters,  there is a possibility that $\{a\in X|f_1(a)=\cdots=f_p(a)=0\}\neq \{O\}$. As our aim is to construct algorithms for studying the system $F$ on $X$, it is necessary, beforehand,  to take away these values of parameters that constitute  constructible sets from the parameter space for computing local cohomology. 

Here, we describe an algorithm for decomposing $\bar{K}^m$ into  ${\cal  S}=\{\A_1,\ldots,\A_k\}$ and ${\cal D}=\{\A_{k+1},\ldots,\A_l\}$ where $\langle F \rangle$ is zero-dimensional on $\A_i$ and nonzero-dimensional on $\A_j$ in a polynomial ring, for $1\le i\le k, k+1\le j \le l$. This decomposition is possible by mainly computing a comprehensive Gr\"obner system of $F$. We adopt the following definition of comprehensive Gr\"obner systems, because this definition is suitable to compute dimensions of ideals in the algorithm {\bf ZeroDimension}. (The following  definition is different from the original one). 

For any $g \in R[x]$  and $GP \subset R[x]$, $\het(g)$ (resp. $\hm(g), \hc(g)$, $\mdeg(g)$) is the head term (resp. the head monomial, the head coefficient, the multidegree) of a polynomial $g$ so that $\hm(g)=\hc(g)\cdot \het(g)$ and $\het(g)=x^{\mdeg(g)}$ hold and $\het(GP)=\{\het(g)|g\in GP\}$ where $R$ is $K, K[t]$ or $K(t)$.

\begin{defn}[Comprehensive Gr\"obner system (CGS)]\label{cgs}\normalfont
Let fix a term order. Let $F$ be a subset of $(K[t])[x]$, 
${\A}_1,\ldots, {\A}_l$ strata in 
${\bar{K}}^m$ and $GP_1,\ldots,GP_l$  subsets of $(K[t])[x]$.  
A finite set ${\cal G}=\Set{({\A}_1,GP_1),\ldots, ({\A}_l, GP_l)}$ of 
pairs is called a {\bf comprehensive Gr\"obner system (CGS)} on ${\A}_1 \cup \cdots \cup {\A}_l$ for $F$ 
if $\sigma_{\bar{a}}(GP_i)$ is a Gr\"obner basis of the ideal 
$\langle \sigma_{\bar{a}}(F) \rangle$ in $\bar{K}[x]$ and $\langle \het(\sigma_{\bar{a}}(GP_i))\rangle=\langle \het(GP_i) \rangle$ 
for each $i=1,\ldots,l$ and $\bar{a} \in {\A}_i$. 
Each $({\A}_i, GP_i)$ is called a {\bf segment} of ${\cal G}$. We simply say ${\cal G}$ is a comprehensive Gr\"obner system for $F$ if ${\A}_1 \cup \cdots \cup {\A}_l={\bar{K}}^m$. 
\end{defn}

 After obtaining a CGS of $F$ w.r.t a total degree term order, as each segment of the CGS has the property $\langle \het(\sigma_{\bar{a}}(GP_i))\rangle=\langle \het(GP_i) \rangle$, the dimension of $\langle GP_i \rangle$ is easily decided in $\bar{K}[x]$. Since an algorithm for computing a CGS terminates, the following algorithm clearly terminates.\\

\noindent
\SEN \vspace{1mm}
\noindent 
{\bf Algorithm 1.} {\bf (ZeroDimension)}\vspace{-3mm}\\
\SEN \vspace{1mm}
\noindent 
{\bf Specification: ZeroDimension(}$F${\bf )}\\
 Testing dimensions of a parametric ideal $\langle F \rangle$ on $\bar{K}^m$.\\
{\bf Input:} $F$: a set of parametric polynomials in $(K[t])[x]$\\
{\bf Output:} $({\cal S},{\cal D})$: ${\cal S}=\{(\A_1,GP_1)\ldots,(\A_k,GP_k)\}$ is a CGS on $\A_1\cup\cdots\cup\A_k$ for $F$ s.t. for all $\bar{a} \in \A_i$, $\langle \sigma_{\bar{a}}(F)\rangle$ is zero-dimensional in $\bar{K}[x]$, for each $i=1,..,k$. 
${\cal D}=\{(\A_{k+1},GP_{k+1})\ldots,$ $(\A_l,$ $GP_l)\}$ is a CGS on $\A_{k+1}\cup\cdots\cup\A_{l}$ for $F$  such that for all $\bar{b} \in \A_j$, $\langle \sigma_{\bar{b}}(F)\rangle$ is not zero-dimensional in $\bar{K}[x]$, for each $j=k+1,\ldots,l$. 
$\bar{K}^m=\A_1\cup \cdots \cup\A_k\cup \A_{k+1}\cup \cdots\cup \A_l$.\\
{\bf BEGIN}\\
${\cal S}\gets \emptyset$; \ \ ${\cal D} \gets \emptyset$ ; \ 
$C \gets$ compute a CGS on $\bar{K}^m$ of $F$ w.r.t. a total degree term order\\
{\bf while} $C\neq \emptyset$ {\bf do} \\
\hspace*{2mm}select $(\A,GP)$ from $C$; $C\gets C\backslash \{(\A,GP)\}$; $d \gets $compute the dimension of $\langle GP \rangle$ in $K[x]$\\
\hspace*{2mm}{\bf if} $d=0$ {\bf then} \ ${\cal S}\gets {\cal S}\cup{\{(\A,GP)\}}$ \ {\bf else} \ ${\cal D}\gets {\cal D}\cup{\{(\A,GP)\}}$ \ {\bf end-if}\\
{\bf end-while}\\
{\bf return}$({\cal S},{\cal D})$ \\
{\bf END} \vspace{-3mm} \\
\noindent
\SEN \vspace{-3mm}\ \\

In our implementation, we adopt Nabeshima's algorithm \citep{Na12} for computing comprehensive Gr\"obner systems, because the algorithm is much more useful than others for computing dimensions of parametric ideals. 

\begin{defn}\normalfont 
Using the same notation as in the above algorithm, let $({\cal S},{\cal D})$ be an output of {\bf ZeroDimension}$(F)$. Then, for each $i=1\ldots, k$, $\A_i$ is called a {\bf safe} stratum, and for each $j=k+1\ldots, l$, $\A_j$ is called a {\bf danger} stratum.
\end{defn}

\begin{exmp}\label{dim}\normalfont 
Let $f=x_1^4+tx_1^2x_2^2+x_2^4$ be a polynomial with a parameter $t$ in $(\C[t])[x_1,x_2]$. A CGS of $F=\{\frac{\partial f}{\partial x_1}, \frac{\partial f}{\partial x_2}\}$ w.r.t. the total degree reverse lexicographic term order s.t. $x_1 \prec x_2$, is $\{(\V(t),\{x_1^3,x_2^3\}),(\V(t-2),\{x_1^2x_2+x_2^3,x_1^3+x_1x_2^2\}),(\V(t+2),\{x_1^2x_2-x_2^3,x_1^3-x_1x_2^2\}),(\C\backslash \V(t(t^2-4)),\{tx_1^2x_2+2x_2^3,2x_1^3+tx_1x_2^2,(t^2-4)x_1x_2^3,(t^2-4)x_2^5\})\}$. 

 If the parameter $t$ belongs to $\V(t)$ or $\C\backslash \V(t(t^2-4))$, then $\langle \frac{\partial f}{\partial x}, \frac{\partial f}{\partial y} \rangle$ is zero-dimensional. If the parameter $t$ belongs to $\V(t-2)$ or $\V(t-2)$, then $\langle \frac{\partial f}{\partial x}, \frac{\partial f}{\partial y} \rangle$ is one-dimensional. Therefore, ${\cal S}=\{(\V(t),\{x_1^3,x_3^3\}),(\C\backslash \V(t(t^2-4)),\{ax_1^2x_2+2x_2^3,2x_1^3+tx_1x_2^2,(t^2-4)x_1x_2^3,(t^2-4)x_2^5\})\}$ and ${\cal D}=\{(\V(t-2),\{x_1^2x_2+x_2^3,x_1^3+x_1x_2^2\}),(\V(t+2),\{x_1^2x_2-x_2^3,x_1^3-x_1x_2^2\})\}$. That is, $\V(t), \C\backslash \V(t(t^2-4))$ are safe strata, and $\V(t-2), \V(t+2)$ are danger strata.
\end{exmp}

Let $({\cal S},{\cal D})$ denote an output of {\bf ZeroDimension}$(F)$ where ${\cal S}=\{(\A_1,$ $GP_1)\ldots,$ $(\A_k,GP_k)\}$ and ${\cal D}=\{(\A_{k+1},GP_{k+1})\ldots,(\A_l,GP_l)\}$ (notation is from the algorithm {\bf ZeroDimension}). Since for all $\bar{a}\in \A_1\cup \cdots \cup\A_k$, $\langle \sigma_{\bar{a}}(F)\rangle$ is zero-dimensional in $\bar{K}[x]$, $\langle \sigma_{\bar{a}}(F)\rangle$ is also zero-dimensional in $\bar{K}[[x]]$. However, in general, for all $\bar{b}\in \A_{k+1}\cup \cdots \cup\A_l$, it is NOT possible for us to say that $\langle \sigma_{\bar{b}}(F)\rangle$ is not zero-dimensional in $\bar{K}[[x]]$. For some  $\bar{b}\in \A_{k+1}\cup \cdots \cup\A_l$, $\langle \sigma_{\bar{b}}(F)\rangle$ may be zero-dimensional in $\bar{K}[[x]]$.

After decomposing the parameter space $\bar{K}^m$ into safe strata and danger strata by the algorithm {\bf ZeroDimension}, we compute bases of the vector space $H_F$ on safe strata and danger strata, separately.  Actually, this decomposition lets us construct an efficient algorithm for computing the bases. (See section~3.3).\\

As the set $F$ has parameters, the structure of the vector spaces $H_F$ depends on the values of parameters $t$. Here, we introduce a definition of parametric local cohomology system of $H_F$.

\begin{defn}\label{55}\normalfont 
Using the same notation as in the above, let ${\A}_i, \B_j$ strata in $\bar{K}^m$ and $S_i$ a subset of $(K[t]_{\A_i})[\xi]$ where $1\le i\le l$ and $1\le j \le k$. Set ${\cal S}=\{({\A}_1,S_1),\ldots,({\A}_l, S_l)\}$ and ${\cal D}=\{\B_1,\ldots,\B_k\}$. Then, a pair $({\cal S}, {\cal D})$ is called a {\bf parametric local cohomology system} of $H_F$ on $\A_1\cup \cdots \cup \A_l\cup \B_1\cup\cdots\cup \B_k$, if for all $i\in\{1,\ldots,l\}$ and $\bar{a} \in {\A}_i$, $\sigma_{\bar{a}}(S_i)$ is a basis of the vector space $H_{\sigma_{\bar{a}}(F)}$, and for all $j\in\{1,\ldots,k\}$ and $\bar{b} \in {\B}_j$, $\{c\in X|\sigma_{\bar{b}}(f_1)(c)=\cdots=\sigma_{\bar{b}}(f_p)(c)=0\}\neq\{O\}$ where 
$H_{\sigma_{\bar{a}}(F)}:=\{\psi \in \bar{K}[\xi] \mid \sigma_{\bar{a}}(f_1)\ast \psi=\sigma_{\bar{a}}(f_2)\ast \psi$ $=\cdots=\sigma_{\bar{a}}(f_p)\ast \psi=0\}.$
\end{defn}

After here, we represent ``a parametric local cohomology system of $H_F$ on $\bar{K}^m$'' as simply ``$H_F$'' which is the abbreviation. Similarly, we call ``a  parametric local cohomology system of $H_F$ on a stratum $\A$'' {\bf `` bases of (the vector space) $H_F$ on $\A$''}.\\

As this section 3 presents thirteen algorithms for computing bases of the vector space $H_F$, Fig. 1 illustrates the relations of the all algorithms. The main algorithm is {\bf ALCohomolog}. 

\begin{figure}[ht]
\begin{center}
\includegraphics[scale=0.5]{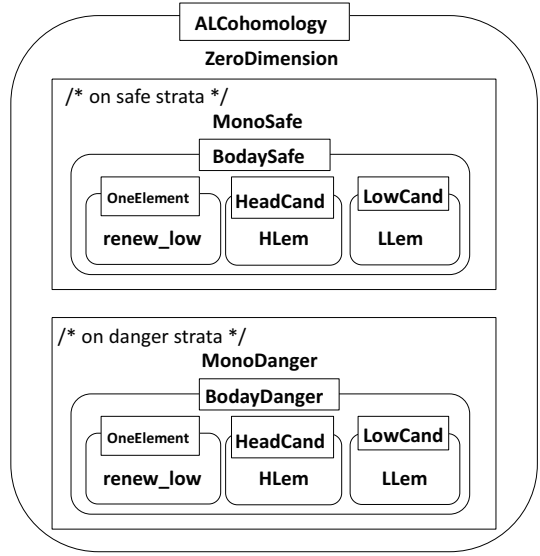}
\caption{relations of all algorithms}
\end{center}
\end{figure}

First, we introduce in section 3.2 an algorithm for computing bases of the vector space $H_F$ on safe strata. Second, we describe in section 3.3 an algorithm for computing bases of the vector space $H_F$ on danger strata.

\subsection{Computation of algebraic local cohomology with parameters on safe strata}
Here, we present an algorithm for computing bases of algebraic local cohomology classes $H_F$, on safe strata. This section consists of three parts. In section 3.2.1, an algorithm for computing monomial elements of bases of $H_F$ is introduced. In section 3.2.2 and 3.2.3, an algorithm for treating with elements, which form linear combination $(\sum c_\lambda \xi^\lambda)$, of bases of $H_F$, is given.


\subsubsection{Monomial elements}
Here, we give an algorithm for computing monomial elements of bases of $H_F$. 
Before describing the algorithm, we define some notation.
\begin{notation}\normalfont 
Let $GP$ be a set of polynomials in  $(K[t])[x]$ and $g \in GP$.
\begin{enumerate}
\item[(1)]The set of monomials of $g$ is denoted by $\Mono(g)$, i.e., 
$\Mono(g):=\{a_\lambda x^\lambda |g=\sum_{\lambda \in \N^n}a_\lambda x^\lambda \text{ where } a_\lambda  \in K[t] \text{ and } a_\lambda \neq 0\}.$ Moreover, the set of monomials of the set $GP$ is denoted by $\Mono(GP)$, i.e., 
$\Mono(GP):=\bigcup_{g\in GP}\Mono(g)$.
\item[(2)]For all $i \in \{1,\ldots,n\}$, a map $\CV$ is defined as {\Large {\it c}}hanging {\Large {\it v}}ariables $x_i$ into $\xi_i$. The inverse map $\CV^{-1}$ is defined as changing variables $\xi_i$ into $x_i$. That is, for any $g\in (K[t])[x]$, $\CV (g)$ is in $(K[t])[\xi]$. The set $\CV(GP)$ is also defined as $\CV(GP)=\{\CV(g)|g\in GP\}$.
\end{enumerate}
\end{notation}

For instance, for $\frac{2}{3}x_1^2x_2+5x_1, 3x_1^2+4x_2 \in K[x_1,x_2]$, then $\CV(\frac{2}{3}x_1^2x_2+5x_1)=\frac{2}{3}\xi_1^2\xi_2+5\xi_1$ and $\CV(\{\frac{2}{3}x_1^2x_2+5x_1, 3x_1^2+4x_2\})=\{\frac{2}{3}\xi_1^2\xi_2+5\xi_1, 3\xi_1^2+4\xi_2\}$ in $K[\xi_1,\xi_2]$ where variables $\xi_1,\xi_2$ are corresponding to variables $x_1, x_2$.

\begin{prop}\normalfont 
Let $(\mathcal{S},\mathcal{D})$ be an output of ${\bf ZeroDimension}(F)$ and $(\A, GP)\in {\cal S}$. Assume that $B$ is a CGS of the monomial ideal $\langle \CV(\Mono(GP))\rangle$ in $(K(t))[\xi]$ on $\A$, and $(\A',G')\in B$. Then, a monic monomial $\psi=\xi_1^{\alpha_1}\xi_2^{\alpha_2}\cdots\xi_n^{\alpha_n}$ which does not belong to $\langle \het(G')\rangle$, has the property $f_i\ast \psi=0$ for each $1\le i \le p$. Namely, all terms which do not belong to $\langle \het(G')\rangle$, are members of bases of $H_F$ on $\A'$.\vspace{-3.0mm}
\end{prop}
\begin{pf}
Let $ \psi = \xi_1^{\alpha_1}\xi_2^{\alpha_2}\cdots \xi_n^{\alpha_n} $ be a monomial s.t. $ \psi \notin \langle ht(G^{\prime})\rangle  $. 
By Definition~\ref{cgs}, for all $\bar{a}\in \A'$, $\langle \het(G') \rangle=\langle \het(\sigma_{\bar{a}}(G'))\rangle=\langle \CV(\Mono($ $\sigma_{\bar{a}}(GP)))\rangle$. As $\langle \het(G') \rangle$ is a zero-dimensional ideal and $\psi \notin \langle \het(G') \rangle$, for all $\xi_1^{\lambda_1}\xi_2^{\lambda_2}\cdots\xi_n^{\lambda_n}\in \CV($ $\Mono(F))$, there always exists $j\in \{1,2,\ldots, n\}$ such that $\lambda_j>\alpha_j$ where $(\alpha_1,\alpha_2,\ldots,\alpha_n),$ $(\lambda_1,\lambda_2,\ldots,\lambda_n) \in \N^n$. Therefore, by the multiplication, $f_i\ast \psi=0$.
\end{pf}
This proposition gives rise to the following algorithm to compute monomial elements of bases of $H_F$ on $\A$. Since the termination of Nabeshima's algorithm \citep{Na12} is guaranteed, the following algorithm terminates.\\

\noindent
\SEN \vspace{1mm}
\noindent 
{\bf Algorithm 2.} {\bf (MonoSafe)}\vspace{-3mm}\\
\SEN \vspace{1mm}
\noindent 
{\bf Specification: MonoSafe(}$\A,GP${\bf )}\\
 Computing monomial elements of bases of $H_F$ on a safe stratum $\A$. \\
{\bf Input:} $(\A,GP)$: a segment of a CGS of $F$ such that for all $\bar{a} \in \A$, $\langle \sigma_{\bar{a}}(F)\rangle$ is zero-dimensional in $\bar{K}[x]$. (This is from ZeroDimension($F$).) \\
{\bf Output:} ${\cal M}$ : a finite set of triples $(\A',M,G)$ such that the set $M$ includes all monomial elements of bases of $H_F$ on $\A'$, and the elements of $M$ do not belong to $\langle G \rangle$.\\
{\bf BEGIN}\\
${\cal M}\gets \emptyset$ ;  \ $B \gets$ compute a CGS of $\CV(\Mono(GP))$ on $\A$\\
{\bf while} $B\neq \emptyset$ {\bf do} \\
\hspace*{4mm}select $(\A',G')$ from $B$; \ $B\gets B\backslash \{(\A',G')\}$; \ $G\gets\het(G')$\\
\hspace*{4mm}$M\gets $ compute monomial elements which do not belong to $\langle G \rangle$ in $K[\xi]$ \ \ $(\ast 1)$\\
\hspace*{4mm}${\cal M} \gets {\cal M} \cup \{(\A',M,G)\}$ \\
{\bf end-while}\\
{\bf return} \ ${\cal M}$ \\
{\bf END}\vspace{-3mm} \\
\SEN \vspace{-3mm} \ \\

Let us remark that as $\langle GP\rangle$ is a zero-dimensional ideal on $\A$, the set $M$ consists of finitely many monomial elements. Note that monomial elements, on danger strata, will be considered in section 3.3. 

We illustrate the algorithm {\bf MonoSafe} with the following example.
\begin{exmp}\label{mono}\normalfont 
Let $f=x_1^4+tx_1^2x_2^2+x_2^4$ be a polynomial with a parameter $t$ in $(\C[t])[x_1,x_2]$. Set $F=\{\frac{\partial f}{\partial x_1}, \frac{\partial f}{\partial x_2}\}$. Then, $F$ satisfies generically $\{a\in X| \frac{\partial f}{\partial x_1}(a)=\frac{\partial f}{\partial x_2}(a)=0\}=\{O\}$ where $X$ is a neighborhood of the origin $O$ of $\C^2$. From Example~\ref{dim}, $(\V(t),\{x_1^3,x_2^3\})$ and $(\C\backslash \V(t(t^2-4)),\{tx_1^2x_2+2x_2^3,2x_1^3+tx_1x_2^2,(t^2-4)x_1x_2^3,(t^2-4)x_2^5\})$ can be inputs of the algorithm {\bf MonoSafe}. 
\begin{enumerate}
\item[(1)] Take $(\V(t),\{x_1^3,x_2^3\})$ as an input of the algorithm {\bf MonoSafe}. Then, a CGS of $\CV(\{x_1^3,x_2^3\})$ on $\V(t)$, is $\{(\V(t),\{\xi_1^3,\xi_2^3\})\}$. Set $G_1=\het(\{ \xi_1^3,\xi_2^3 \})=\{\xi_1^3,\xi_2^3\}$. Then, all elements of $M_1=\{1,\xi_1,\xi_2,\xi_1^2,\xi_1\xi_2,\xi_2^2,\xi_1^2\xi_2,\xi_1\xi_2^2,\xi_1^2\xi_2^2\}$ do not belong to $\langle G_1\rangle$. See $\bullet$ in Fig. 2. $M_1$ can be a subset of bases of $H_F$ on $\V(t)$.  

\item[(2)]Take $(\C\backslash \V(t(t^2-4)), P)$ where $GP=\{tx_1^2x_2+2x_2^3,2x_1^3+tx_1x_2^2,(t^2-4)x_1x_2^3,(t^2-4)x_2^5\}$. As $\Mono(GP)=\{tx_1^2x_2,2x_2^3,2x_1^3,tx_1x_2^2,(t^2-4)x_1x_2^3,(t^2$ $-4)x_2^5\}$, a CGS of $\langle \CV(\Mono(GP)) \rangle$ on $\C\backslash \V(t(t^2-4))$ is $\{\C\backslash \V(t(t^2-4)),\{\xi_1^3,t\xi_1^2\xi_2,t\xi_1\xi_2^2,\xi_2^3\})\}$. Set $G_2=\het(\{\xi_1^3,t\xi_1^2\xi_2,t\xi_1\xi_2^2,\xi_2^3\})=\{\xi_1^3,\xi_1^2\xi_2,$ $\xi_1\xi_2^2,\xi_1^3\}$ and compute monomial elements which do not belong to $\langle G_2\rangle$. Then, we obtain $M_2=\{1,\xi_1,\xi_2,\xi_1^2,\xi_1\xi_2,$ $\xi_2^2\}$ which can be a subset of bases of $H_F$ on $\C\backslash \V(t(t^2-4))$. See $\bullet$ in Fig. 3.

\end{enumerate}
 \begin{minipage}{.60\linewidth}
 \ \ \ \ \ \ \ \ \ \ \ \ \ 
\setlength\unitlength{1truecm}
\begin{picture}(4.5,3.5)(0,0)

\put(0,0){\vector(1,0){4}}
\put(0,0){\line(0,1){1}}
\put(0,0){\line(1,0){4}}
\put(0,1.5){\line(1,0){1.5}}

\put(1.5,1.5){\line(0,-1){1.5}}
\put(0,0){\vector(0,1){3}}
\put(-0.3,-0.3){$(0,0)$}
\put(0,0){$\bullet $}
\put(0,0.5){$\bullet $}
\put(0,1){$\bullet $}

\put(0.05,1.62){$\xi_2^3$}

\put(0.5,0){$\bullet $}
\put(0.5,0.5){$\bullet $}
\put(0.5,1){$\bullet $}

\put(1,0){$\bullet $}
\put(1,0.5){$\bullet $}
\put(1,1){$\bullet $}
\put(1.55,0.12){$\xi_1^3$}

\put(-0.35,2.65){$\xi_2$}
\put(4,-0.3){$\xi_1$}
\put(1.7,-0.9){\text{Fig. 2}}
\put(1.0,-0.5){\text{exponents of } $M_1$}

\end{picture}
\end{minipage}
\hspace{2mm}
\begin{minipage}{12.6\baselineskip}

\setlength\unitlength{1truecm}
\begin{picture}(4.5,3.5)(0,0)
\put(0,0){\vector(1,0){4}}
\put(0,0){\line(0,1){1}}
\put(0,1.5){\line(1,0){0.5}}
\put(1.5,0){\line(0,1){0.5}}
\put(1,0.5){\line(1,0){0.5}}
\put(1.05,0.62){$\xi_1^2\xi_2$}
\put(1,0.5){\line(0,1){0.5}}
\put(0.5,1){\line(1,0){0.5}}

\put(0.5,1.5){\line(0,-1){0.5}}
\put(0,0){\vector(0,1){3}}
\put(-0.3,-0.3){$(0,0)$}
\put(0,0){$\bullet $}
\put(0,0.5){$\bullet $}
\put(0,1){$\bullet $}

\put(0.05,1.62){$\xi_2^3$}

\put(0.5,0){$\bullet $}
\put(0.5,0.5){$\bullet $}

\put(1,0){$\bullet $}
\put(1.55,0.12){$\xi_1^3$}
\put(0.55,1.12){$\xi_1\xi_2^2$}

\put(-0.35,2.6){$\xi_2$}
\put(4,-0.3){$\xi_1$}
\put(1.7,-0.9){\text{Fig. 3}}
\put(1.0,-0.5){\text{exponents of } $M_2$}

\end{picture}

\end{minipage}
\vspace{5.0mm}
 \ \\
\end{exmp}

\subsubsection{Head terms of linear combination elements and the main algorithm}
Here, we illustrate an algorithm for computing bases of $H_F$. Before describing the algorithm, first we treat with elements, which form linear combination $(\sum c_\lambda \xi^\lambda)$, of bases of $H_F$. Especially, we discuss how to decide head terms of the linear combination elements $(\sum c_\lambda \xi^\lambda)$.
Second, an algorithm for computing bases of $H_F$ on safe strata, is given. Note that an algorithm for deciding lower terms, will be described in section 3.2.3.
 
Let us recall the following lemma which follows from the fact that if $\psi\in H_{F}$, so is $x_i\ast \psi \in H_{F}$ for each $i=1,\ldots, n$. This lemma informs us candidates of head terms in $H_{F}$.

\begin{lem}[\cite{TN09}]\label{lemma3}\normalfont 
Let $\Lambda_{F}$ denote the set of exponents of head terms in $H_{F}$ and $\lambda=(\lambda_1,\ldots,\lambda_n)\in \mathbb{N}^n$. Let $\Lambda_{F}^{(\lambda)}$ denote a subset of $\Lambda_{F}$ : \ 
$\Lambda_{F}=\{\lambda \in \N^n| \exists \psi \in H_{F}  \text{ s.t.} \het(\psi)=\xi^\lambda\} \text{ and }  \Lambda_{F}^{(\lambda)}=\{\lambda'\in \Lambda_{F} |\lambda'\prec \lambda\}.$
If $\lambda \in \Lambda_{F}$, then, for each $j=1,2,\ldots ,n, (\lambda_1,\lambda_2,\ldots,\lambda_{j-1},\lambda_j-1,\lambda_{j+1},\ldots,\lambda_n)$ is in $\Lambda^{(\lambda)}_{F}$, provided $\lambda_j \ge 1$.
\end{lem}

Let $\xi^\lambda$ be a term  where $\lambda=(\lambda_1,\ldots,\lambda_n)\in \N^n$. We call $\xi^{\lambda}\cdot \xi_i$ a {\bf neighbor} of $\xi^{\lambda}$ for each $i=1,\ldots,n$. Then, $|\xi^\lambda \cdot \xi_i|=\sum_{i=1}^{n}\lambda_i+1$.

\begin{notation}\normalfont 
Let $T$ be a set of terms in $K[\xi]$. Then, we define the neighbor of $T$ as {\bf Neighbor}$(T)$, i.e., {\bf Neighbor}$(T):=\{\tau\cdot \xi_i|\tau\in T, i=1,\ldots, n\}$.

\end{notation}

The following corollary is a direct consequence of Lemma~\ref{lemma3}.
\begin{cor}\label{coro1}\normalfont 
Let $\text{TList}^{(d)}=\{\xi^{\lambda}|\lambda \in \Lambda_{F}, |\xi^\lambda|=d \}$. If for all $i \in \{1,\ldots, n\}$, $\tau=\xi_1^{\lambda_1}\cdots \xi_n^{\lambda_n} \in \text{TList}^{(d+1)}$ satisfies $\xi_i|\tau$, 
then, $\{\tau/\xi_i| \lambda_i\neq 0, i\in \{1,\ldots, n\}\}\subset \text{TList}^{(d)}$ where $(\lambda_1,\ldots,\lambda_n)\in \N^n$.
\end{cor}

Let $T$ be a subset of $\text{TList}^{(d)}$. 
Then, by Corollary~\ref{coro1}, there is a possibility that an element of {\bf Neighbor}$(T)$ belongs to $\text{TList}^{(d+1)}$. This fact makes up the following algorithm which outputs new candidates for head terms. \\

\noindent
\SEN \vspace{1mm}
\noindent 
{\bf Algorithm 3.} {\bf (HLem)}\vspace{-3mm}\\
\SEN \vspace{1mm}
\noindent 
{\bf Specification: HLem(}$T, \text{TList}^{(d)}${\bf )}\\
 Making new candidates for head terms from $T$.\\
{\bf Input:} $T$: a set of terms whose total degree are $d$, and $T \subseteq \text{TList}^{(d)}.$\\
{\bf Output:} S: a set of new candidates whose total degree are $d+1$.\\
{\bf BEGIN}\\
S$\gets \emptyset$; \ B$\gets$ {\bf Neighbor}$(T)$\\
{\bf while} B $\neq \emptyset$ {\bf do} \\
\hspace*{4mm}select $\tau$ from B; \ B$\gets$ B$\backslash \{\tau\}$\\
\hspace*{4mm}{\bf for} $i$ from $1$ to $n$ {\bf do} \ Flag$\gets 1$\\
\hspace*{8mm}{\bf if} $\xi_i|\tau$ {\bf then}\\
\hspace*{12mm}{\bf if} $\tau/\xi_i\notin \TList^{(d)}$ {\bf then} Flag$\gets 0$ ; \ {\bf break} \ {\bf end-if} \\ 
\hspace*{8mm}{\bf end-if}\\
\hspace*{4mm}{\bf end-for}\\
\hspace*{8mm}{\bf if} Flag$=1$ {\bf then} \ \ S$\gets$ S$\cup\{\tau\}$ \ {\bf end-if}\\ 
{\bf end-while}\\
{\bf return} \ S \\
{\bf END}\vspace{-3mm} \\
\SEN \vspace{-3mm} \ \\

If $\tau$ is not in the set of head terms of $H_F$ (written $\het(H_F)$), then neighbors of $\tau$ are not in $\het(H_F)$. This means that if $\tau$ is not in $\het(H_F)$, then it is unnecessary to compute elements which are divided by $\tau$, as candidates for head terms. This fact makes up the following notation {\bf NonMember}. We also give the notation {\bf Car} and {\bf Cdr} which are exploited in the some algorithms.
\begin{notation}\label{td}\normalfont 
Let $T$ be a set of terms in $K[\xi]$ and $\tau$ be the smallest element in $T$ w.r.t. the term order (of Definition~\ref{monoorder}).
\begin{enumerate}
\item[(1)] Let $\FL$ be a set of terms in $K[\xi]$ such that for all $\xi^\lambda \in \FL$, $\lambda$ is not in $\Lambda_F$ where $\Lambda_{F}$ is the set of exponents of head terms in $H_F$. Then, the notation {\bf NonMember} of $(T,\FL)$ is defined by 
$\text{{\bf NonMember}}(T,\FL):=\{\psi\in T|\varphi \nmid \psi \text{ for all }\varphi\in \FL\}.$
\item[(2)] The notation {\bf Car} and {\bf Cdr} for $T$, are defined as follows
$$\text{\bf Car}(T):=\SetDef{\psi\in T}{|\psi|=|\tau|}, \ \ \ \ \text{\bf Cdr}(T):=T\backslash \text{{\bf Car}}(T).$$
\item[(3)]Suppose that $T^{(d)}=\{\xi^\lambda\in T| |\xi^\lambda|=d\in \N\}$ and $\TT=\{T^{(d_1)},T^{(d_2)},\ldots,$ $T^{(d_u)}\}$ where $d_i\in \N$ and $T^{(d_i)}\neq \emptyset$ for each $i=1,\ldots,u$. Let $d_j$ be the minimal number in $\{d_1,\ldots,d_u\}$. Then, the same notation {\bf Car} and {\bf Cdr} are defined, for a set of sets $\TT$, as follows \vspace{-3.0mm}
$$\text{\bf Car}(\TT):=T^{(d_j)}, \ \ \ \ \text{\bf Cdr}(\TT):=\TT\backslash \text{{\bf Car}}(\TT).$$

\end{enumerate}
\end{notation}

Let ${\cal M}$ be an output of {\bf MonoSafe}$(\A, GP)$ where $(\A, GP)$ is a segment of a CGS of $F$. Suppose that $(\A',M',G')$ is an element of ${\cal M}$. Remember that all elements of $M'$ do not belong to $\langle G' \rangle$. Since clearly $G' \subset \text{{\bf Neighbor}}(M')$, elements of $G'$ become candidates of head terms in $H_F$. The use of this property makes candidates of the head terms, efficiently.  
\begin{cor}\normalfont 
Using the same notation as in the above discussion, Notation~\ref{td} and Lemma~\ref{lemma3}, let $M^{(d)}=\{\xi^{\lambda}|\xi^\lambda \in M', \ |\xi^\lambda|=d \}$ and $T^{(d)}=\text{TList}^{(d)}\backslash M^{(d)}$. Then, elements of $\text{{\bf Neighbor}}(T^{(d)})$ and $G'$ can be candidates of head terms in $H_F$. 
\end{cor}

Suppose that $\GList=\{G^{(d_1)},\ldots,G^{(d_u)}\}$ and $\FL$ is a set of terms in $K[\xi]$ such that for all $\xi^\lambda \in \FL$, $\lambda$ is not in $\Lambda_F$ where $G^{(d_j)}=\{\xi^{\gamma} \in G'||\xi^{\gamma}|=d_j\}$ and $j\in \{1,\ldots, u\}$. 
Now, we introduce how to obtain a set of candidates of head terms in $H_F$ from $T^{(d)}$ and $\GList$. In order to make the set of the candidates, the following four cases are considered. \vspace{3.0mm}

\begin{center}
{\bf Case (i)} $T^{(d)}=\emptyset \wedge \GList=\emptyset$. \ {\bf Case (ii)} $T^{(d)}=\emptyset \wedge \GList\neq \emptyset$. \\
{\bf Case (iii)} $T^{(d)}\neq \emptyset \wedge \GList=\emptyset$. \ 
{\bf Case (iv)} $T^{(d)}\neq \emptyset \wedge \GList\neq \emptyset$.\vspace{3.0mm}
\end{center}

In case (i), our main algorithm terminates, because any candidates of the head terms can not be made by the sets. In case (ii), {\bf Car}($\GList$) has to be considered as a set of the next candidates w.r.t. the term order. In case (iii), {\bf NonMember}({\bf HLem}($T^{(d)},$ $\TList^{(d)}$), $\FL$) has to be considered as a set of the next candidates whose total degree is $d+1$. In case (iv), for any $\tau \in${\bf Car}($\GList$), if $|\tau|-d=1$, then {\bf NonMember}({\bf HLem}($T^{(d)},\TList^{(d)}$), $\FL$)$\cup${\bf Car}($\GList$) has to be considered as a set of the next candidates, otherwise the next candidates is {\bf NonMember}({\bf HLem}($T^{(d)},\TList^{(d)}$), $\FL$).

Let us remark that the algorithm {\bf BodySafe} decides head terms of bases of $H_F$, from bottom to up with respect to the term order (total degree lexicographic  term order). Therefore, the sets $T^{(d)}$ and $\text{TList}^{(d)}$ are already obtained when the following algorithm makes the set of the candidates whose total degree are $d+1$.\\

\noindent
\SEN \vspace{1mm}
\noindent 
{{\bf Algorithm 4.} {\bf (HeadCand)}}\vspace{-3mm}\\
\SEN \vspace{1mm}
\noindent 
{\bf Specification: HeadCand(}$T^{(d)}, \GList, \TList^{(d)}, \FL${\bf )}\\
 Making new candidates for head terms. \\
{\bf Input:} $T^{(d)}$, $\GList$, $\TList^{(d)}$, $\FL$: described above.\\
{\bf Output:} $\CT$: a set of new candidates for head terms (or {\bf Car}$(\GList)$). $\GList$: renewed $\GList$; \ $T^{(d)}$:renewed $T^{(d)}$. \\
{\bf BEGIN}\\
{\bf if} $T^{(d)}=\emptyset \wedge \GList=\emptyset$ {\bf then} \ \ \verb|/* case (i) */|\\
\hspace*{4mm}$\CT\gets \emptyset$ ; \ {\bf return}$(\CT,\GList,T^{(d)})$\\
{\bf else if} $T^{(d)}=\emptyset \wedge \GList\neq \emptyset$ {\bf then}  \ \ \verb|/* case (ii) */|\\
\hspace*{4mm}$\CT\gets\text{{\bf Car}}(\GList)$; \ $\GList\gets\text{{\bf Cdr}}(\GList$) ; \ {\bf return}$(\CT,\GList,T^{(d)})$\\
{\bf else if} $T^{(d)}\neq \emptyset \wedge \GList= \emptyset$ {\bf then}  \ \ \verb|/* case (iii) */|\\
\hspace*{4mm}$\CT\gets$ {\bf NonMember}({\bf HLem}($T^{(d)},\TList^{(d)}$), $\FL$); \ $d\gets d+1$; \ $T^{(d)}\gets \emptyset$ \\
\hspace*{4mm}{\bf return}$(\CT,\GList,T^{(d)})$\\
{\bf else if} $T^{(d)}\neq \emptyset \wedge \GList\neq \emptyset$ {\bf then} \ \ \verb|/* case (iv) */|\\
\hspace*{4mm}$G\gets${\bf Car}($\GList$)\ ; \ GL$\gets\text{{\bf Cdr}}(\GList$); select $\xi^\gamma$ from $G$ \\
\hspace*{4mm}{\bf if} $|\xi^\gamma|-d>1$ {\bf then}\\
\hspace*{4mm}$\CT\gets$ {\bf NonMember}({\bf HLem}($T^{(d)},\TList^{(d)}$), $\FL$); \ $d\gets d+1$ ; \ $T^{(d)}\gets \emptyset$ \\ 
\hspace*{4mm}{\bf return}$(\CT,\GList,T^{(d)})$\\
\hspace*{4mm}{\bf end-if}\\
\hspace*{4mm}{\bf if} $|\xi^\gamma|-d=1$ {\bf then} \\
\hspace*{4mm}$\CT\gets${\bf NonMember}({\bf HLem}($T^{(d)},\TList^{(d)}$), $\FL$)$\cup G$ ; \ $\GList \gets$GL\\
\hspace*{4mm}{\bf return}$(\CT,\GList,T^{(d)})$\\
\hspace*{4mm}{\bf end-if}\\
{\bf end-if}\\
{\bf END}\vspace{-3mm} \\
\SEN \vspace{-3mm} \ \\

The algorithm {\bf BodySafe} consists of mainly two parts, computing candidates for head terms and lower terms. For each part, the algorithm makes use of sets as intermediate data. As this is a dynamic algorithm, each intermediate data is often renewed in the algorithm. As sets $\SList, \MList, \LList, \GList, \CT, T^{(d)}, \CL$ are frequently used in algorithms on a stratum, we fix the meaning of the sets as follows.

\begin{notation}\label{nota1}\normalfont 
$\SList:=\{\psi \in K(t)[\xi]| \ \psi \text{ is a linear combination element of a basis}\}$. \\
$\MList:=\{\psi \in K[\xi]| \ \psi  \text{ is a monic monomial element of a basis}\}$.  \\
$\LList:=\{\xi^{\lambda}\in K[\xi]| \ \xi^{\lambda} \text{ is a lower term of }\psi \text{ where }\psi\in \SList\}.$\\
$\CT:=\{\tau \in K[\xi]|\ \tau\text{ is a candidate for head terms of a basis}\}$.\\
$\FL:=\{\tau\in K[\xi]|\ \tau\text{ is a failed candidate for head terms}\}$.\\
$\GList:=\bigcup_{i}\{\{\xi^{\gamma}\in G|\ |\xi^{\gamma}|=d_i\}\}$ described in the algorithm {\bf HeadCand}. \\
$T^{(d)}:=\{\tau \in K[\xi]|\ \tau\text{ is a head term whose total degree is }d \}$.\\
$\CL:=\{\xi^{\lambda}\in K[\xi]|\ \xi^{\lambda}\text{ is a candidate for lower terms for some } \tau \in \CT\}$.
\end{notation}

As $F$ has parameters, when we compute bases of $H_F$ by the main algorithm {\bf ALCohomology}, the parameter space $\bar{K}^m$ is decomposed to suitable strata for the bases. Hence, on each stratum, the sets above are decided. Note that when the algorithm terminates, then a set $\SList \cup \MList$ becomes a basis of $H_F$ on each stratum.

In the following two algorithms, sets $\EL, \LL, \UU, \RR$ are used for algorithmic consistency, to decide lower terms. The sets will be explained in section 3.2.3.

The main algorithm {\bf ALCohomology} consists of two parts for safe strata and danger strata. The first part an algorithm {\bf BodySafe} for safe strata, is given in this section. The second part an algorithm {\bf BodyDanger} for danger strata will be discussed in section~3.3. 

Suppose that $\mathcal{Q}$ is a list. Then, $\mathcal{Q}[i]$ means the $i$th element of the list $\mathcal{Q}$. For example, let $\mathcal{Q}=[\A,\CT,\GList]$, then $\mathcal{Q}[1]=\A, \mathcal{Q}[2]=\CT$ and $\mathcal{Q}[3]=\GList$. 
In the following algorithms, lists $\mathcal{Q}$, $\mathcal{E}$ and $\MList^{(d)}=\{\tau \in \MList||\tau|=d\}$ play actively. \\
\ \\
\noindent
\SEN \vspace{1mm}
\noindent 
{\bf Algorithm 5.} {\bf (ALCohomology)}\vspace{-3mm}\\
\SEN \vspace{1mm}
\noindent 
{\bf Specification: ALCohomology(}$F, k${\bf )}\\
 Computing bases of a vector space $H_F$ with parameters. \\
{\bf Input:} $F=\{f_1,\ldots,f_p\}$: $F\subset (K[t])[x]$ satisfying {\bf generically} $\{a\in X|$ $f_1(a)=\cdots=f_p(a)=0\}=\{O\}$ where $X$ is a neighborhood of the origin $O$ of $K^n$. $\nu\in \N$: an estimated bound of dimensions of the vector space $H_F$ or a sufficient big number (see section 3.3). \\
{\bf Output:} $(\mathcal{S},\mathcal{D})$: $\mathcal{S}$ is a set of lists $[\A,\SList,\MList,\LList, \FL]$ where $\SList\cup \MList$ is a basis of $H_F$ on $\A$, $\LList$ is a set of lower terms of $\SList$ and $\FL$ is a set of failed candidates for head terms on $\A$.\\
{\bf BEGIN}\\
$\CT\gets \emptyset$;\ $\SList\gets \emptyset$; \ $\LList\gets \emptyset$;\ $\FL\gets \emptyset$;\ $\LL\gets \emptyset$; \ $\RR\gets \emptyset$; \ $\EL\gets \emptyset$\\
$\UU\gets \emptyset$; \ $\mathcal{AC}\gets \emptyset$; \ $\mathcal{DL}\gets \emptyset$; \ $(\mathcal{Z},\mathcal{N})\gets \text{{\bf ZeroDimension}}(F)$
\renewcommand{\arraystretch}{1} \\
\begin{tabular}{|l|}\hline
\verb|/*on safe strata */|\\
{\bf while}$\mathcal{Z}\neq \emptyset$ {\bf do}\\
select $Z_1$ from $\mathcal{Z}$; \ $\mathcal{Z}\gets \mathcal{Z}\backslash \{Z_1\}$; \ $\mathcal{M}\gets{\bf MonoSafe}(Z_1)$ \\
\hspace*{2mm}{\bf while} $\mathcal{M}\neq \emptyset$ {\bf do} \\
\hspace*{2mm}select $(\A,\MList,G)$ from $\mathcal{M}$; \ $\mathcal{M} \gets \mathcal{M}\backslash \{(\A,\MList,G)\}$ \\
\hspace*{2mm}${\GList\gets \bigcup_{i}\{\{\xi^{\gamma}\in G||\xi^{\gamma}|=d_i\}\}}$; \ $\tau\gets$ the smallest element in $G$ w.r.t. $\prec$; \ $d\gets |\tau|$\\
\hspace*{2mm}$T^{(d)}\gets \emptyset$; \ $\mathcal{Q}\gets [\A,\CT,\GList,T^{(d)},\SList,\MList,\LList, \FL,\LL,\EL,\RR,\UU]$ \\
\hspace*{2mm}$\mathcal{AC}\gets\mathcal{AC}\cup\{\mathcal{Q}\}$\\
\hspace*{2mm}{\bf end-while}\\
{\bf end-while}\\
$\text{Coho}\gets{\bf BodySafe}(\mathcal{AC},F)$ \\\hline
\end{tabular} 
 \\ 
\begin{tabular}{|l|}\hline
\verb|/*on danger strata */|\\
{\bf while}$\mathcal{N}\neq \emptyset$ {\bf do}\\
select $N_1$ from $\mathcal{N}$; \ $\mathcal{N}\gets \mathcal{N}\backslash \{N_1\}$; \ $\mathcal{M}\gets{\bf MonoDanger}(N_1)$\\
\hspace*{2mm}{\bf while} $\mathcal{M}\neq \emptyset$ {\bf do} \\
\hspace*{2mm}select $(\A,\MList,G)$ from $\mathcal{M}$; \ $\mathcal{M} \gets \mathcal{M}\backslash \{(\A,\MList,G)\}$ \\
\hspace*{2mm}{\bf if} $\MList\neq \emptyset$ {\bf then} \\
\hspace*{2mm}${\GList\gets \bigcup_{i}\{\{\xi^{\gamma} \in G||\xi^{\gamma}|=d_i\}\}}$; \ $\tau\gets$ the smallest element in $G$ w.r.t. $\prec$; \ $d\gets |\tau|$ \\
\hspace*{2mm}$T^{(d)}\gets \emptyset$; \ $\mathcal{Q}\gets [\A,\CT,\GList,T^{(d)},\SList,\MList,\LList, \FL,\LL,\EL,\RR,\UU]$ \\
\hspace*{2mm}$\mathcal{DL}\gets\mathcal{DL}\cup\{\mathcal{Q}\}$\\
\hspace*{2mm}{\bf else} \ $\mathcal{D}\gets \mathcal{D}\cup\{\A\}$ \\
\hspace*{2mm}{\bf end-if} \\
\hspace*{2mm}{\bf end-while}\\
{\bf end-while}\\
$(\text{Co},\mathcal{D}_1)\gets{\bf BodyDanger}(\nu,\mathcal{DL},F)$\\\hline
\end{tabular}\\
\renewcommand{\arraystretch}{1}
$\mathcal{S}\gets \text{Coho}\cup \text{Co}$; \ $\mathcal{D}\gets \mathcal{D}\cup\mathcal{D}_1$ \\
{\bf return} \ $(\mathcal{S},\mathcal{D})$ \\
{\bf END}\vspace{-3mm} \\
\SEN \vspace{-2mm} \ \\

\noindent
\SEN \vspace{1mm}
\noindent 
{\bf Algorithm 6.} {\bf (BodySafe)}\vspace{-3mm}\\
\SEN \vspace{1mm}
\noindent 
{\bf Specification: BodySafe(}$\mathcal{AC}$,F{\bf )}\\
Computing bases of algebraic local cohomology $H_F$ for $\mathcal{AC}$.\\
{\bf Input:} $\mathcal{AC}$: a set of lists $([\A,\CT,\GList,T^{(d)},\SList,\MList,\LList, \FL,\LL,\EL,$ $\RR,\UU])$.\\
{\bf Output:} $\mathcal{S}$: a set of lists $[\A,\SList,\MList,\LList, \FL]$ where $\SList\cup \MList$ is a basis of $H_F$ on $\A$ , $\LList$ is a set of lower terms of $\SList$, and $\FL$ is a set of failed candidates for head terms on $\A$.\\
{\bf BEGIN}\\
$\mathcal{S} \gets \emptyset$\\
{\bf while} $\mathcal{AC}\neq \emptyset$ {\bf do}\\
select $\mathcal{E}=[\A,\CT,\GList,T^{(d)},\SList,\MList,\LList, \FL,\LL,\EL, \RR,\UU]$ from $\mathcal{AC}$\\ 
$\mathcal{AC} \gets \mathcal{AC} \backslash\{\mathcal{E}\}$ \renewcommand{\arraystretch}{1}\\
\begin{tabular}{|l|}\hline
\hspace*{4mm}{\bf if} $\CT\neq \emptyset$ {\bf then} \ $\xi^\gamma\gets \text{{\bf Car}}(\CT)$; \ $\CT \gets\text{{\bf Cdr}}(\CT)$\\
\hspace*{4mm}{\bf else}  \\
\hspace*{4mm}$(\CT,\GList,T^{(d)})\gets  \text{{\bf HeadCand}}(T^{(d)}, \GList, \MList^{(d)}\cup T^{(d)},\FL)$ ($\diamondsuit$1)\\
\hspace*{8mm}{\bf if} $\CT\neq \emptyset$ {\bf then}\\
\hspace*{8mm}$\xi^\gamma\gets \text{{\bf Car}}(\CT)$; \ $\CT \gets\text{{\bf Cdr}}(\CT)$\\
\hspace*{8mm}{\bf else}\\
\hspace*{8mm}$\mathcal{S}\gets\mathcal{S}\cup\{[\A,\SList,\MList,\LList, \FL]\}$\\
\hspace*{8mm}{\bf end-if}\\
\hspace*{4mm}{\bf end-if} \\\hline\hline
\hspace{-1.0mm}$(\CL,\UU,\EL)\gets\text{{\bf LowCand}}(\xi^\gamma,\SList,\MList,\LList,\LL,\UU,\RR,\EL)$ ($\diamondsuit$2)\\\hline\hline
$\mathcal{Q}\gets[\CT,\GList,\MList,\UU]$\\
$\mathcal{P}\gets\text{{\bf OneElement}}(\xi^\gamma,\CL,\A,T^{(d)},\EL,\FL,\SList,\LList,\mathcal{Q},F)$ \ \ \ \ \ \ \ \ \ \ \ ($\diamondsuit$3)\\\hline
\end{tabular} 
\renewcommand{\arraystretch}{1}\\
$\mathcal{S}\gets\mathcal{S}\cup \text{{\bf BodySafe}}(\mathcal{P})$\\
{\bf end-while}\\
{\bf return} $\mathcal{S}$\\
{\bf END}\vspace{-3mm} \\
\SEN \vspace{-3mm} \ \\


The algorithm {\bf BodySafe} consists of three parts ($\diamondsuit$1), ($\diamondsuit$2) and ($\diamondsuit$3). In ($\diamondsuit$1), new candidates for head terms are computed. The part ($\diamondsuit$1) was already described in the beginning of this section. In ($\diamondsuit$2), candidates ($\CL$) of $\xi^\gamma$'s lower terms are computed. The part ($\diamondsuit$2) will be described in section 3.3. Here, we do not explain the part ($\diamondsuit$2), but by seeing the operation of $\CL$, one can understand the flow of the algorithm {\bf BodySafe}. In ($\diamondsuit$3), an element $\xi^{\gamma}+\sum_{\lambda\in \CL}c_{\lambda}\xi^{\lambda}$ ($c_{\lambda}\in K(t))$ is tested whether it can be in $H_F$ or not. That is, linear combination elements are decided in the part ($\diamondsuit$3). Note that in ($\diamondsuit$3), a list $\mathcal{Q}$ is not essentially used by the algorithm {\bf OneElement}. The list $\mathcal{Q}$ is just used in order to shorten the algorithm. The part ($\diamondsuit$3) is given as follows.\\

\noindent
\SEN \vspace{1mm}
\noindent 
{\bf Algorithm 7.} {\bf OneElement}\vspace{-3mm}\\
\SEN \vspace{1mm}
\noindent 
{\bf Specification: OneElement(}$\xi^\gamma,\CL,\A,T^{(d)},\EL,\FL,\SList,\LList,\mathcal{Q},\{f_1,\ldots,f_p\}${\bf )}\\
 Testing whether $\xi^{\gamma}+\sum_{\lambda\in \CL}c_{\lambda}\xi^{\lambda}$ is in $H_F$ or not.\\
{\bf Input:} $\xi^\gamma,\CL,\A,T^{(d)},\EL,\FL,\SList,\LList,\mathcal{Q}$: described in the algorithm {\bf BodySafe}.\\
{\bf Output:} $\mathcal{L}$: a set of lists $[\A,\CT,\GList,T^{(d)},\SList,\MList,\LList, \FL,\LL,\EL,$ $\RR,\UU]$.\\
{\bf BEGIN}\\
$\mathcal{L}\gets \emptyset$; \ $E\gets \emptyset$\\
$\psi\gets$ set $ \xi^\gamma+\sum_{\xi^\lambda\in \CL}c_{\lambda}\xi^{\lambda}$ where $c_{\lambda}$'s are indeterminates\\
{\bf for} $i$ {\bf from} 1 {\bf to} $p$ {\bf do}\\
\hspace*{4mm}$\psi \gets f_i\ast \psi$ \ \ \ \ \ \ \ \ \verb|/*check| $f_i\ast \psi=0$. $f_i\ast \psi\in (K[t,c_\lambda])[\xi]$\verb|*/| \\
\hspace*{4mm}{\bf while} $\psi\neq 0$ {\bf do}\\
\hspace*{8mm}$E\gets E\cup \{\hc(\psi)=0\}$; \ $\psi\gets \psi-\hm(\psi)$\\
\hspace*{4mm}{\bf end-while}\\
{\bf end-for}\renewcommand{\arraystretch}{1}\\
\begin{tabular}{|l|}\hline
($\mathcal{A}_1,\mathcal{A}_2)\gets$solve the system $E$ of parametric linear equations on $\A$.\\\hline
\end{tabular} \ $(\ast 1)$\renewcommand{\arraystretch}{1}\\
{\bf while} $\mathcal{A}_1\neq \emptyset$ {\bf do}\\
\hspace*{4mm}select an element $(\A',[c_{\lambda}$'s solutions$])$ from $\mathcal{A}_1$; \ $\mathcal{A}_1\gets \mathcal{A}_1\backslash \{(\A',[c_{\lambda}$'s solutions$])\}$ \\
\hspace*{4mm}$\psi'\gets $ substitute $c_{\lambda}$'s solutions into $\psi$; \ $\SList\gets \SList\cup \{\psi'\}$; \ $T^{(d)}\gets T^{(d)}\cup \{\xi^{\gamma}\}$ \\
\hspace*{4mm}$(\EL,\LL,\RR,\LList)\gets$ {\bf renew\_low}$(1,\EL,\psi'-\xi^\gamma,\LList)$\\
\hspace*{4mm}$\mathcal{L}\gets\mathcal{L}\cup\{[\A',\mathcal{Q}[1],\mathcal{Q}[2],T^{(d)},\SList,\mathcal{Q}[3],\LList,\FL,\LL,\EL,\RR,\mathcal{Q}[4]]\}$\\
{\bf while-end}\\
{\bf while} $\mathcal{A}_2\neq \emptyset$ {\bf do}\\
\hspace*{4mm}select an element $\A'$ from $\mathcal{A}_2$; \ $\mathcal{A}_2\gets \mathcal{A}_2\backslash \{\A'\}$; \ $\FL\gets \FL\cup\{\xi^\gamma\}$\\
\hspace*{4mm}$(\EL,\LL,\RR,\LList)\gets$ {\bf renew\_low}$(0,\EL,\xi^\gamma,\LList)$\\
\hspace*{4mm}$\mathcal{L}\gets\mathcal{L}\cup\{[\A',\mathcal{Q}[1],\mathcal{Q}[2],T^{(d)},\SList,\mathcal{Q}[3],\LList,\FL,\LL,\EL,\RR,\mathcal{Q}[4]]\}$\\
{\bf while-end}\\
{\bf return} $\mathcal{L}$\\
{\bf END}\vspace{-3mm} \\
\SEN \vspace{-3mm} \ \\

If $\psi=\xi^\gamma+\sum_{\xi^\lambda\in \CL}c_{\lambda}\xi^{\lambda}$ is in $H_F$, then $\psi$ satisfies conditions $f_i\ast \psi=0$ for each $i=1,\ldots, p$. These conditions give us a set $E$ of $c_{\lambda}$'s linear equations. Thus, by solving the system $E$, we know whether $\psi$ is in $H_F$ or not. Namely, if solutions of $c_{\lambda}$'s exist, then $\psi\in H_F$, and if the solutions of $c_{\lambda}$'s do not exist, then $\psi\notin H_F$. 

Let us remark that as the system of equations $E$ has parameters, the stratum $\A$ has to be decomposed into suitable strata for the solutions. For instance, let $t$ be a parameter and $x,y$ be variables. Consider a system $``tx+y=4, \ 3x+2y=-9"$ of parametric linear equations on $\C\backslash \V(t)$. Then, the system has the following solutions; if the parameter $t$ belongs to $\C\backslash \V(t(3t+4)(2t-3))$, then $x=\frac{17}{2t-3}, y=\frac{-9t-12}{2t-3}$, if the parameter $t$ belongs to $\V(3t+4)$, then $x=-3, y=0$, and if a parameter $t$ belongs to $\V(2t-3)$, then $E$ has no solution. There exist several algorithms for solving a system of parametric linear equations \citep{G1,sit92}. In our implementation, we extend the Gaussian elimination method to handle parametric cases.

In the box $(\ast 1)$ of the algorithm {\bf OneElement}, $\mathcal{A}_1$ means a set of pairs $(\A',[c_{\lambda}$'s solutions$])$ and $\mathcal{A}_2$ means a set of strata such that for any stratum of $\mathcal{A}_2$, the system has no solution. The algorithm {\bf OneElement} has a subalgorithm {\bf renew\_low} which computes candidates of lower terms of $\xi^\gamma$ and is given in section 3.2.3.
\begin{thm}\normalfont 
The first part of the algorithm {\bf ALCohomology} (i.e., {\bf BodySafe}) terminates and outputs a set $\text{Coho}$ which has a list $[\A,\SList,\MList,$  $\LList, \FL]$ such that $\SList\cup \MList$ is a basis of $H_F$ on $\A$. 
\end{thm}
\begin{pf}
The algorithms {\bf LowCand} and {\bf renew\_low} are considered in section 3.2.3. and the termination and correctness are discussed in section 3.2.3. 
In the algorithm {\bf ZeroDimension}, the parameter space $\bar{K}^m$ is decomposed to a finite number of strata. As we described, in the algorithm {\bf OneElement}, an algorithm for solving the system of parametric equations, outputs a finite number of strata \citep{G1,sit92}. Since the algorithm {\bf BodySafe} works on safe strata, $\langle F \rangle$ is zero-dimensional on the strata. This means that $H_F$ is a finite-dimensional vector space \citep{TN09, TNN09}. Therefore, the first part of the algorithm {\bf ALCohomology} (i.e., {\bf BodySafe}) generates a finite number of strata. Thus, the algorithm terminates. Moreover, clearly  all elements of $\SList\cup \MList$ are linearly independent on $\A$,  $\SList\cup \MList$ is a basis of $H_F$ on $\A$.
\end{pf}

\begin{exmp}\label{linear}\normalfont 
Let $f=x_1^4+tx_1^2x_2^2+x_2^4\in (\C[t])[x_1,x_2]$. Set $F=\{\frac{\partial f}{\partial x_1}, \frac{\partial f}{\partial x_2}\}$. Then, $F$ satisfies generically $\{a\in X| \frac{\partial f}{\partial x_1}(a)=\frac{\partial f}{\partial x_2}(a)=0\}=\{O\}$ where $X$ is a neighborhood of the origin $O$ of $\C^2$.  The term order is the total degree lexicographic term order such that $\xi_1\prec \xi_2$.

\begin{enumerate}
\item[(0)] $\CT=\emptyset$,  $\SList=\emptyset$, $\FL=\emptyset$.
\end{enumerate}
{\bf Case (1):} From Example~\ref{mono}, the output of {\bf MonoSafe}$(\V(t),\{x_1^3,x_2^3\})$ is $(\V(t),$ $\MList,G_1)$ where $\MList=\{1,\xi_1,\xi_2,\xi_1^2,\xi_1\xi_2,\xi_2^2,\xi_1^2\xi_2,\xi_1\xi_2^2,\xi_1^2\xi_2^2\}$ and $G_1=\{\xi_1^3,$ $\xi_2^3\}$. Then, $\GList=\{\{\xi_1^3,\xi_2^3\}\}$ and the smallest element in $G_1$ is $\xi_1^3$. Since $|\xi_1^3|=3$, set $T^{(3)}=\emptyset$.
\begin{enumerate}
\item[(1-1)] As $T^{(3)}=\emptyset$, $\GList=\{\{\xi_1^3,\xi_2^3\}\}$ and $\CT=\emptyset$, $\CT$ and $\GList$ are renewed as $\CT=\text{{\bf Car}}(\GList)=\{\xi_1^3,\xi_2^3\}$ and $\GList=\emptyset$. So, $\text{{\bf Car}}(\CT)=\xi_1^3$ and $\CT=\text{{\bf Cdr}}(\CT)=\{\xi_2^3\}$. Then, the algorithm \text{{\bf LowCand}} outputs the empty set as the set of candidates of $\xi_1^3$'s lower terms. This means that there is no candidate for lower terms of $\xi_1^3$. Since $\frac{\partial f}{\partial x_1}\ast \xi_1^3=4, \frac{\partial f}{\partial x_2}\ast \xi_1^3=0$, we obtain equations $4=0$ and $0=0$. Clearly, $4=0$ is false. Hence, $\xi_1^3$ can not be a member of $H_F$. Renew $\FL$ as $\{\xi_1^3\}$.

\item[(1-2)]As $\CT=\{\xi_2^3\}$, the new candidate for head terms is $\xi_2^3$. Renew $\CT$ as $\text{{\bf Cdr}}(\CT)=\emptyset$. Then, the algorithm \text{{\bf LowCand}} outputs $\{\xi_1^3\}$ as the set of candidates for lower terms. Set $\xi_2^3+c_{(3,0)}\xi_1^3$ where  $c_{(3,0)}$ is an indeterminate. Since $\frac{\partial f}{\partial x_1}\ast (\xi_2^3+c_{(3,0)}\xi_1^3)=4c_{(3,0)}, \ \frac{\partial f}{\partial x_2}\ast (\xi_2^3+c_{(3,0)}\xi_1^3)=4$, we obtain equations $4c_{(3,0)}=0$ and $4=0$. Clearly, $4=0$ is false. Hence, $\xi_2^3$ can not be a head term in $H_F$. Renew $\FL$ as $\{\xi_1^3, \xi_2^3\}$. This process terminates, because $\CT=\GList=\emptyset$.
\end{enumerate}
{\bf Case (2):} From Example~\ref{mono}, the output of {\bf MonoSafe}$(\C\backslash \V(t(t^2-4)),\{tx_1^2x_2$ $+2x_2^3,2x_1^3+tx_1x_2^2,(t^2-4)x_1x_2^3,(t^2-4)x_2^5\})$, is $(\C\backslash \V(t(t^2-4)),\MList,G_2)$ where $\MList=\{1,\xi_1,\xi_2,\xi_1^2,$ $\xi_1\xi_2,$ $\xi_2^2\}$ and 
$G_2=\{\xi_1^3,\xi_1^2\xi_2,\xi_1\xi_2^2,\xi_2^3\}$. Then, 
$\GList=\{\{\xi_1^3,\xi_1^2\xi_2,$ $\xi_1\xi_2^2,\xi_2^3\}\}$ and the smallest element in $G_2$ is $\xi_1^3$. Since $|\xi_1^3|=3$, set $T^{(3)}=\emptyset$.
\begin{enumerate}
\item[(2-1)]As $T^{(3)}=\emptyset$, $\GList\neq \emptyset$ and $\CT=\emptyset$,  $\CT$ and $\GList$ are renewed as 
$\CT=\text{{\bf Car}}(\GList)=\{\xi_1^3,\xi_1^2\xi_2,\xi_1\xi_2^2,\xi_2^3\}$ and $\GList=\emptyset$. So, $\text{{\bf Car}}(\CT)=\xi_1^3$ and 
$\CT=\text{{\bf Cdr}}(\CT)=\{\xi_1^2\xi_2,\xi_1\xi_2^2,\xi_2^3\}$. By the same reasoning as in (1-1), $\xi_1^3$ can not be a member of $H_F$. Renew $\FL$ as $\{\xi_1^3\}$.
\item[(2-2)]As $\CT=\{\xi_1\xi_2^2, \xi_1^2\xi_2,\xi_2^3\}$, the new candidate of a head term in $H_F$, is $\xi_1^2\xi_2$. Renew $\CT$ as $\text{{\bf Cdr}}(\CT)=\{\xi_2^3,\xi_1\xi_2^2\}$. Then, the algorithm \text{{\bf LowCand}} outputs $\{\xi_1^3\}$ as the set of candidates of $\xi_1^2\xi_2$'s lower terms. Set $\xi_1^2\xi_2+c_{(3,0)}\xi_1^3$ where  $c_{(3,0)}$ is an indeterminate. Since $\frac{\partial f}{\partial x_1}\ast (\xi_1^2\xi_2+c_{(3,0)}\xi_1^3)=4c_{(3,0)}, \frac{\partial f}{\partial x_2}\ast (\xi_1^2\xi_2+c_{(3,0}\xi_1^3)=2t$, a system of equations is `` $4c_{(3,0)}=0$,  $2t=0$". As we work on the stratum $\C\backslash \V(t(t^2-4))$, $2t=0$ is false. Hence, $\xi_1^2\xi_2$ can not be a head term in $H_F$. $\FL=\{\xi_1^3\}\cup\{ \xi_1^2\xi_2\}=\{\xi_1^3, \xi_1^2\xi_2\}.$

\item[(2-3)]As $\CT=\{\xi_1\xi_2^2,\xi_2^3\}$, the next candidate is $\xi_1\xi_2^2$. Renew $\CT$ as $\text{{\bf Cdr}}(\CT)=\{\xi_2^3\}$. Then, the algorithm \text{{\bf LowCand}} outputs $\{\xi_1^3, \xi_1^2\xi_2\}$ as the set of candidates of $\xi_1\xi_2^2$'s lower terms. Set $\psi=\xi_1\xi_2^2+c_{(2,1)}\xi_1^2\xi_2+c_{(3,0)}\xi_1^3$ where  $c_{(2,1)},c_{(3,0)}$ are indeterminates. Since $\frac{\partial f}{\partial x_1}\ast \psi=4c_{(3,0)}+2t, \frac{\partial f}{\partial x_2}\ast \psi=2tc_{(2,1)}$, a system of equations is ``$4c_{(3,0)}+2t=0$, $2tc_{(2,1)}=0$". Solve the linear equations, then $c_{(3,0)}=-\frac{1}{2}t$ and $c_{(2,1)}=0$. Hence, $\xi_1\xi_2^2-\frac{1}{2}t\xi_1^3$ is a member of a basis of $H_F$ on $\C\backslash \V(t(t^2-4))$. Thus, $\SList=\{\xi_1\xi_2^2-\frac{1}{2}t\xi_1^3\}$, $T^{(3)}=\{\xi_1\xi_2\}$.

\item[(2-4)]As $\CT=\{\xi_2^3\}$, the next candidate is $\xi_2^3$. Renew $\CT$ as $\text{{\bf Cdr}}(\CT)=\emptyset$. Then, the algorithm \text{{\bf LowCand}} outputs $\{\xi_1^3, \xi_1^2\xi_2\}$ as the set of candidates of $\xi_2^3$'s lower terms. Set $\psi=\xi_2^3+c_{(2,1)}\xi_1^2\xi_2+c_{(3,0)}\xi_1^3$ where  $c_{(2,1)},c_{(3,0)}$ are indeterminates. Since $\frac{\partial f}{\partial x_1}\ast \psi=4c_{(3,0)}, \frac{\partial f}{\partial x_2}\ast \psi=2tc_{(2,1)}+4$, a system of equations is ``$4c_{(3,0)}=0$,  $2tc_{(2,1)}+4=0$". Solve the linear equations, then $c_{(3,0)}=0$ and $c_{(2,1)}=-\frac{1}{2t}$. Hence, $\xi_2^3-\frac{1}{2t}\xi_1^2\xi_2$ is a member of a basis of $H_F$ on $\C\backslash \V(t(t^2-4))$. Thus, $\SList=\{\xi_1\xi_2^2-\frac{1}{2}t\xi_1^3, \xi_2^3-\frac{1}{2t}\xi_1^2\xi_2\}$, $T^{(3)}=\{\xi_1\xi_2^2, \xi_2^3\}$. 
Now, as $\CT=\emptyset$, the set $\CT$ should be renewed by the algorithm {\bf HeadCand}. Since {\bf Neighbor}$(T^{(3)})=\{\xi_1^2\xi_2^2, \xi_1\xi_2^3,\xi_2^4\}$ and $\xi_1^2\xi_2| \xi_1^2\xi_2^2$ where $\xi_1^2\xi_2$ is in $\FL$, $\xi_1^2\xi_2^2$ can not be in $\CT$. Therefore, the renewed $\CT$ is $\{\xi_1\xi_2^3,\xi_2^4\}$.

\item[(2-5)]As $\CT=\{\xi_1\xi_2^3,\xi_2^4\}$, the next candidate is $\xi_1\xi_2^3$. Renew $\CT$ as $\text{{\bf Cdr}}(\CT)=\{\xi_2^4\}$. Then, the algorithm \text{{\bf LowCand}} outputs $\{\xi_1^3, \xi_1^2\xi_2, \xi_1^3\xi_2,$ $ \xi_1^2\xi_2^2, \xi_1^4\}$ as the set of candidates of $\xi_1\xi_2^3$'s lower terms. Set $\psi=\xi_1\xi_2^3+c_{(2,2)}\xi_1^2\xi_2^2+c_{(3,1)}\xi_1^3\xi_2+c_{(4,0)}\xi_1^4+c_{(2,1)}\xi_1^2\xi_2+c_{(3,0)}\xi_1^3$ where $c_{(2,2)},c_{(3,1)}, c_{(4,0)}, c_{(2,1)},c_{(3,0)}$ are indeterminates. $\frac{\partial f}{\partial x_1}\ast \psi=(4c_{(4,0)}+2tc_{(2,2)})\xi_1+(4c_{(3,1)}+2t)\xi_2+4c_{(3,0)}, \ \frac{\partial f}{\partial x_2}\ast \psi=(2tc_{(3,1)}+4)\xi_1+2tc_{(2,2)}\xi_2+2tc_{(2,1)}$. If $\psi$ is in $H_F$, then  $\psi$ satisfies the conditions $\frac{\partial f}{\partial x_1}\ast p=0$ and $\frac{\partial f}{\partial x_2}\ast p=0$. Hence, we have to check the five equations $4c_{(4,0)}+2tc_{(2,2)}=0, \ 4c_{(3,1)}+2t=0, \ 2tc_{(3,1)}+4=0, \ 2tc_{(2,2)}=0, \ 2tc_{(2,1)}=0$ on $\C\backslash \V(t(t^2-4))$. The two equations $4c_{(3,1)}+2t=0$ and $2tc_{(3,1)}+4=0$ hold only if $t=\pm2$. Therefore, $\psi$ can not be in $H_F$. $\FL=\{\xi_1^3, \xi_1^2\xi_2\}\cup\{\xi_1\xi_2^3\}=\{\xi_1^3, \xi_1^2\xi_2,\xi_1\xi_2^3\}.$

\item[(2-6)]The next candidate is $\xi_2^4$ and $\CT=\emptyset$. The algorithm \text{{\bf LowCand}} outputs 
$\{\xi_1^3, \xi_1^2\xi_2,$ $\xi_1^4, \xi_1^3\xi_2, \xi_1^2\xi_2^2, \xi_1\xi_2^3\}$ as the set of candidates of $\xi_2^4$'s lower terms. 
Set $\psi=\xi_2^4+c_{(1,3)}\xi_1\xi_2^3+c_{(2,2)}\xi_1^2\xi_2^2+c_{(3,1)}\xi_1^3\xi_2+c_{(4,0)}\xi_1^4+c_{(2,1)}\xi_1^2\xi_2+c_{(3,0)}\xi_1^3$ 
where $c_{(1,3)}, c_{(3,1)},$ $c_{(4,0)}, c_{(2,1)},c_{(3,0)}$ are indeterminates. $\frac{\partial f}{\partial x_1}\ast \psi=(2tc_{(1,3)}+2tc_{(2,2)}+4c_{(4,0)})\xi_1+4c_{(3,1)}\xi_2+4c_{(3,0)},$ \ 
$ \frac{\partial f}{\partial x_2}\ast \psi=(4c_{(1,3)}+2tc_{(3,1)})\xi+(2tc_{(2,2)}+4)\xi_2+2tc_{(2,1)}$. 
Hence, we have to check the system of equations: $2tc_{(1,3)}+2tc_{(2,2)}+4c_{(4,0)}=0, \ 4c_{(3,1)}=0, \ 4c_{(3,0)}=0, \ 4c_{(1,3)}+2tc_{(3,1)}=0, \ 2tc_{(2,2)}+4=0, \ 2tc_{(2,1)}=0$. Then, the solution is : $c_{(1,3)}=0, c_{(2,2)}=-\frac{1}{t}, c_{(3,1)}=0, c_{(4,0)}=1, c_{(2,1)}=0, c_{(3,0)}=0$. Hence, $\xi_2^4-\frac{1}{t}\xi_1^2\xi_2^2+\xi_1^4$ is a member of a basis of $H_F$ on $\C\backslash \V(t(t^2-4))$. Thus, $\SList=\{\xi_1\xi_2^2-\frac{1}{2}t\xi_1^3, \xi_2^3-\frac{1}{2t}\xi_1^2\xi_2, \xi_2^4-\frac{1}{t}\xi_1^2\xi_2^2+\xi_1^4\}$, $T^{(4)}=\{\xi_2^4\}$. 
As $\CT=\emptyset$, $\CT$ can be renewed as $\{\xi_2^5\}$.
\item[(2-7)]The next candidate is $\xi_2^5$ and $\CT=\emptyset$. The algorithm \text{{\bf LowCand}} outputs 
$\{\xi_1^3, \xi_1^2\xi_2,$ $\xi_1^4, \xi_1^3\xi_2, \xi_1^2\xi_2^2, \xi_1\xi_2^3,\xi_1^5\}$
 as the set of candidates of $\xi_2^5$'s lower terms. Set 
$\psi=\xi_2^5+c_{(5,0)}\xi_1^5+c_{(1,3)}\xi_1\xi_2^3+c_{(2,2)}\xi_1^2\xi_2^2+c_{(3,1)}\xi_1^3\xi_2+c_{(4,0)}\xi_1^4+c_{(2,1)}\xi_1^2\xi_2+c_{(3,0)}\xi_1^3$ where $c_{(5,0)},c_{(1,3)}, c_{(2,2)}, c_{(3,1)}, c_{(4,0)},$ $c_{(2,1)}c_{(3,0)}$ are indeterminates. In this case, there is no solution that satisfies the conditions $\frac{\partial f}{\partial x_1}\ast \psi=0$ and $\frac{\partial f}{\partial x_2}\ast \psi=0$. $\FL=\{\xi_1^3, \xi_1^2\xi_2,\xi_1\xi_2^3,\xi_2^5\}.$ As $\CT=\GList=\emptyset$, this process terminates.
\end{enumerate}
We summarize the results as follows:
\begin{enumerate}
\item[if] a parameter $t$ belongs to $\V(t)$ (i.e., $t=0$), $\{1,\xi_1,\xi_2,\xi_1^2,\xi_1\xi_2,\xi_2^2,\xi_1^2\xi_2,\xi_1\xi_2^2,\xi_1^2\xi_2^2\}$ is a basis of $H_F$ (algebraic local cohomology classes),
\item[if] a parameter $t$ belongs to $\C\backslash \V(t(t^2-4))$ (i.e., $t\neq 0,t\neq\pm 2$), then $\{1,\xi_1,\xi_2,\xi_1^2,\xi_1\xi_2,$ $\xi_2^2, \xi_1\xi_2^2-\frac{1}{2}t\xi_1^3, \xi_2^3-\frac{1}{2t}\xi_1^2\xi_2, \xi_2^4-\frac{1}{t}\xi_1^2\xi_2^2+\xi_1^4\}$ is a basis of $H_F$.
\end{enumerate}
In Fig. 4 and 5, we represent an element of $\MList$ as $\bullet$ and an element of $\het(\SList)$ as $\triangle$. Note that on each stratum, a basis of $H_F$ is $\MList\cup \SList$. As the set $\FL$ plays a key role to construct standard bases (see section 4), we specially give the elements of $\FL$ in the Figures.\vspace{-5.0mm}\\
\begin{figure}[ht]

 \begin{minipage}{.60\linewidth}
 \ \ \ \ \ \ \ \ \ \ \ \ \ 
\setlength\unitlength{1truecm}
\begin{picture}(4.5,4.5)(0,0)

\put(0,0){\vector(1,0){4}}
\put(0,0){\line(0,1){1}}
\put(0,0){\line(1,0){4}}
\put(0,1.5){\line(1,0){1.5}}

\put(1.5,1.5){\line(0,-1){1.5}}
\put(0,0){\vector(0,1){4}}
\put(-0.3,-0.3){$(0,0)$}
\put(0,0){$\bullet $}
\put(0,0.5){$\bullet $}
\put(0,1){$\bullet $}

\put(0.05,1.63){$\xi_2^3$}

\put(0.5,0){$\bullet $}
\put(0.5,0.5){$\bullet $}
\put(0.5,1){$\bullet $}

\put(1,0){$\bullet $}
\put(1,0.5){$\bullet $}
\put(1,1){$\bullet $}
\put(1.55,0.13){$\xi_1^3$}

\put(-0.35,3.6){$\xi_2$}
\put(4,-0.3){$\xi_1$}
\put(-0.6,-0.8){Case (1)}
\put(1.7,-0.9){\text{Fig. 4}}
\put(1.0,-0.5){\text{exponents}}
\end{picture}
\end{minipage}
\hspace{2mm}
\begin{minipage}{12.6\baselineskip}

\setlength\unitlength{1truecm}
\begin{picture}(4.5,4.5)(0,0)
\put(0,0){\vector(1,0){4}}
\put(0,0){\line(0,1){1}}
\put(0,0){\line(1,0){4}}
\put(0,2.5){\line(1,0){0.5}}
\put(1.5,0){\line(0,1){0.5}}
\put(1,0.5){\line(1,0){0.5}}
\put(1.05,0.63){$\xi_1^2\xi_2$}
\put(1,0.5){\line(0,1){1}}
\put(0.5,1.5){\line(1,0){0.5}}

\put(0.5,1.5){\line(0,1){1}}
\put(0,0){\vector(0,1){4}}
\put(-0.3,-0.3){$(0,0)$}
\put(0,0){$\bullet $}
\put(0,0.5){$\bullet $}
\put(0,1){$\bullet $}

\put(0.5,0){$\bullet $}
\put(0.5,0.5){$\bullet $}
\put(0.5,1){$\triangle $}
\put(0,1.5){$\triangle $}
\put(0,2){$\triangle $}

\put(1,0){$\bullet $}
\put(1.55,0.13){$\xi_1^3$}
\put(0.05,2.653){$\xi_2^5$}
\put(-0.35,3.6){$\xi_2$}
\put(0.55,1.63){$\xi_1\xi_2^3$}
\put(4,-0.3){$\xi_1$}
\put(-1.2,-0.8){Case (2)}
\put(1.7,-0.9){\text{Fig. 5}}
\put(1.0,-0.5){\text{exponents}}

\end{picture}

\end{minipage}
 \vspace{10.0mm}
\end{figure}

\end{exmp}

\subsubsection{Lower terms of linear combination elements}

The aim of this section is to construct subalgorithms ``{\bf LowCand}'' and ``{\bf renew\_low}'' which are in the algorithms ``{\bf BodySafe}'' and ``{\bf OneElement}''. Here, we discuss how to compute candidates of lower terms. 
The ideal for computing the candidates efficiently, is to use the information of the intermediate data $\SList$, $\MList$, $\LList$, $\FL$. Before describing the algorithms, we introduce the following useful lemma.

\begin{lem}[\cite{TN09}]\label{lemma4}\normalfont 
Using the same notation as in Lemma~\ref{lemma3}, let $\Delta_{F}$ denote the set of exponents of lower terms in $H_{F}$ and $\Delta_{F}^{(\lambda)}$ denote a subset of $\Delta_{F}$ : \ 
$\Delta_{F}^{(\lambda)}=\{\lambda'\in \Delta_{F} |\lambda'\prec \lambda\}.$
If $\lambda \in \Delta_{F}$, then, for each $j=1,2,\ldots ,n, (\lambda_1, \lambda_2, \ldots,$ $\lambda_{j-1}, \lambda_j-1, \lambda_{j+1}, \ldots, \lambda_n)$ is in $\Delta^{(\lambda)}_{F}\cup \Lambda_{F}^{(\lambda)}$, provided $\lambda_j \ge 1$.
\end{lem}

The algorithm {\bf BodySafe} computes linear combination elements of a basis of $H_F$ from bottom to up with respect to the term order. The next corollary shows a relation between  the indeterminate data ``$\SList, \MList, \LList$'' and new candidates of lower terms.

\begin{cor}\label{cor}\normalfont 
Let $\lambda=(\lambda_1,\ldots,\lambda_n)\in \N^n$ and let $\SList, \MList, \LList$ be indeterminate data in the algorithm {\bf BodySafe}. If $\xi^{\lambda}\in\LList$, then, for each $j=1,2,\ldots ,n$, the term $\xi^\lambda/\xi_i$ is in $\het(\SList)\cup\MList\cup\LList$, provided $\lambda_j \ge 1$.
Conversely, an element of $\text{{\bf Neighbor}}(\het(\SList)\cup\MList\cup\LList)$ becomes a candidate for lower terms.
\end{cor}

This corollary leads us to construct the following algorithm which is essentially same as the algorithm {\bf HLem}.\\

\noindent
\SEN \vspace{1mm}
\noindent 
{\bf Algorithm 8.} {\bf (LLem)} \vspace{-3mm}\\
\SEN \vspace{1mm}
\noindent 
{\bf Specification: LLem(}$\text{Ne}, \SList,\MList,\LList${\bf )}\\
 Making candidates for lower terms from \text{Ne}.\\
{\bf Input:}Ne: a set of terms.  \\
{\bf Output:} $S$: a set of new candidates for lower terms. \\
{\bf BEGIN}\\
$S\gets \emptyset$ \\
{\bf while} $\text{Ne} \neq \emptyset$ {\bf do} \\
\hspace*{4mm}select $\tau$ from $\text{Ne}$; \ $\text{Ne}\gets \text{Ne}\backslash \{\tau\}$\\
\hspace*{4mm}{\bf for} $i$ from $1$ to $n$ {\bf do} \ Flag$\gets 1$\\
\hspace*{8mm}{\bf if} $\xi_i|\tau$ {\bf then}\\
\hspace*{12mm}{\bf if} $\tau/\xi_i\notin \het(\SList)\cup\MList\cup\LList$ {\bf then} Flag$\gets 0$; \ {\bf break} \\
\hspace*{12mm}{\bf end-if} \\ 
\hspace*{8mm}{\bf end-if}\\
\hspace*{4mm}{\bf end-for}\\
\hspace*{8mm}{\bf if} Flag$=1$ {\bf then} \ \ $S\gets S\cup\{\tau\}$ \ {\bf end-if}\\ 
{\bf end-while}\\
{\bf return} \ $S$ \\
{\bf END}\vspace{-3mm} \\
\SEN \vspace{-3mm} \ \\

Let us remark that if a lower term is in $\het(\SList)\cup \MList$, then the lower term can be reduced by elements of $\SList\cup \MList$. Namely, $\LList$ obtained, becomes always a part of candidates for lower terms. Thus, a set of the candidates is $$\CL=\{\text{proper new candidates of lower terms}\}\cup \LList.$$ 

As sets $\EL, \LL, \RR, \UU$ are frequently used in the algorithms {\bf LowCand} and {\bf renew\_low} on a stratum, we fix the meaning of the sets as follows.

\begin{notation}\normalfont 
 \ $\EL:=\{\xi^{\lambda}\in K[\xi]|\ \xi^{\lambda} \text{ is a new candidate for lower terms. }\xi^{\lambda}\notin\LList\}$.\\
$\LL:=\{\xi^{\lambda}|\ \xi^{\lambda} \text{ is a proper new lower term which belong to }\EL\}$. \ 
$\RR:=\EL \backslash \LL$.\\
$\UU:=\{\xi^\beta\in \text{{\bf Neighbor}}(\LL)|\ \xi^\gamma \prec \xi^\beta, \text{ for some }\xi^\gamma \in \CT\}$. \\
Note that a set $\EL\cup \LList$ becomes a set of candidates for lower terms.
\end{notation}

As $\LL$ is a set of new lower terms, by Corollary~\ref{cor}, elements of {\bf Neighbor}$(\LL)$ become candidates of lower terms. Furthermore, elements of $\{\xi^\alpha|\xi^\alpha \prec \xi^\gamma, \xi^\alpha \in \UU\}$ also become candidates of lower terms where $\xi^\gamma$ is a candidate of a head term. \\

\noindent
\SEN \vspace{1mm}
\noindent 
{\bf Algorithm 9.} {\bf (LowCand)} \vspace{-3mm}\\
\SEN \vspace{1mm}
\noindent 
{\bf Specification: LowCand(}$\xi^\gamma,\SList,\MList,\LList,\LL,\UU,\RR,\EL${\bf )}\\
Making candidates for lower terms of $\xi^\gamma$. \\
{\bf Input:} $\xi^\gamma,\SList,\MList,\LList,\LL,\UU,\RR,\EL$: described in  {\bf BodySafe}.\\
{\bf Output:} $(\CL,\UU,\EL)$: elements of $\CL$ are candidates for $\xi^\gamma$'s lower terms. $\UU$ is a renewed set. $\EL$ is a renewed set.\\
{\bf BEGIN}\\
$U\gets \{\xi^\alpha|\xi^\alpha \prec \xi^\gamma, \xi^\alpha \in \UU\}$\\
{\bf if} $\LL=\emptyset$ {\bf then} \\
\hspace*{4mm}$\text{LU}\gets \text{{\bf LLem}}(U,\SList,\MList,\LList)$; \ $\UU\gets \UU\backslash U$\\
\hspace*{4mm}$\EL\gets \EL\backslash \text{LU}$; \ $\CL \gets \LList\cup\EL$ \\
\hspace*{4mm}{\bf return}$(\CL,\UU,\EL)$ \\
{\bf else}  \  \verb|/* make| $\EL$, $\UU$ \verb|from| $\RR$, $\LL$ \verb|.*/|\\
\hspace*{4mm}$\UU \gets (\UU\backslash U)\backslash \{\xi^\gamma\}$; \ $\RR \gets U\cup\RR$ \\ 
\hspace*{4mm}$B\gets \{\beta|\ \xi^\gamma \prec \xi^\beta, \beta \in \text{{\bf Neighbor}}(\LL))\}$; \ $\UU \gets B\cup \UU$ \\ 
\hspace*{4mm}$D\gets \text{{\bf LLem}}(\text{{\bf Neighbor}}(\LL)\backslash B,\SList,\MList,\LList)$ \\
\hspace*{4mm}$\EL\gets (D \backslash (D \cap \RR))\cup \RR$; \ $\CL \gets \EL\cup \LList$ \\
\hspace*{4mm}{\bf return}$(\CL,\UU,\EL)$ \\
{\bf end-if}\\
{\bf END}\vspace{-3mm} \\
\SEN \vspace{-3mm} \ \\

Since the algorithm {\bf OneElement} is dynamic, each intermediate data of $\EL$, $\LL$, $\RR$ and $\LList$, is often renewed in the algorithm. If a system of linear equations has solutions (i.e., $Z=1$ in {\bf renew\_low}), then the proper new lower terms appear as $\LL$. If a system of linear equations does not have any solution (i.e., $Z=0$ in {\bf renew\_low}), then the candidate of a head term becomes a candidate of lower terms because the candidate is always in $\text{{\bf Neighbor}}(\het(\SList)\cup\MList\cup\LList)$. This observation makes the algorithm {\bf renew\_low}.\\

\noindent
\SEN \vspace{1mm}
\noindent 
{\bf Algorithm 10.} {\bf (renew\_low)}\vspace{-3mm}\\
\SEN \vspace{1mm}
\noindent 
{\bf Specification: renew\_low(}$Z,\EL, \psi, \LList${\bf )}\\
 Renewing the sets $\EL$, $\LL$ and $\RR$.\\
{\bf Input:} $Z$: 0 or 1. $\psi$: a polynomial.\\
{\bf Output:} $(\EL,\LL,\RR,\LList)$: renewed sets $\EL,\LL,\RR,\LList$.\\
{\bf BEGIN}\\
{\bf if}  $Z=0$  {\bf then} \ $\LL\gets \emptyset$; \ $\EL\gets \EL\cup\{\psi\}$; \ $\RR\gets \emptyset$\\
{\bf else}\ $\LL\gets \het(\Mono(\psi))\cap\EL$; \ $\LList\gets \LList\cup \LL$ \\ 
\hspace*{4mm}{\bf if} $\LL\neq \emptyset$ {\bf then} \ $\RR\gets\EL\backslash \LL$; \ $\EL\gets \emptyset$ \\
\hspace*{4mm}{\bf end-if}\\
{\bf end-if} \\
{\bf return} $(\EL,\LL,\RR,\LList)$ \\
{\bf END}\vspace{-3mm} \\
\SEN \vspace{-3mm} \ \\


\begin{exmp}\normalfont 
Let us consider Example~\ref{linear}, again.  Here, we show the process for computing candidates for lower terms according to the algorithms {\bf LowCand} and {\bf renew\_low}. Since a set of candidates for lower terms is $\CL$, we mainly observe the set $\CL$. 
\begin{enumerate}
\item[(0)]$\LList=\emptyset, \LL=\emptyset, \RR=\emptyset, \EL=\emptyset, \UU=\emptyset$.
\end{enumerate}
{\bf Case (1):} First, we start to discuss how to compute $\CL$ on $\V(t)$. 
\begin{enumerate}
\item[(1-1)] Take $\xi_1^3$ as a candidate of a head term. The algorithm {\bf LowCand} outputs the empty set as $\CL$. By the algorithm {\bf renew\_low}, the set $\EL$ is renewed as $\{\xi_1^3\}$.

\item[(1-2)]Take $\xi_2^3$ as a candidate of a head term. According to the algorithm {\bf LowCand}, $\CL=\EL\cup \LList=\{\xi_1^3\}$. In Example~\ref{linear}, $\xi_2^3$ can not be a head term in $H_F$. Hence, $\EL$ is renewed as $\{\xi_1^3,\xi_2^3\}$. 
\end{enumerate}
{\bf Case (2):} Second, we discuss how to compute the set $\CL$ on $\C^2\backslash \V(t(t^2-4))$. 
\begin{enumerate}
\item[(2-1)]Take $\xi_1^3$ as a candidate of a head term. The algorithm {\bf LowCand} outputs the empty set as $\CL$. By the algorithm {\bf renew\_low}, the set $\EL$ is renewed as $\{\xi_1^3\}$. 

\item[(2-2)]Take $\xi_1^2\xi_2$ as a candidate of a head term. According to the algorithm {\bf LowCand}, $\CL=\EL\cup \LList=\{\xi_1^3\}$. In Example~\ref{linear}, $\xi_2^3$ can not be a head term in $H_F$. Hence, $\EL$ is renewed as $\{\xi_1^3,\xi_1^2\xi_2\}$.

\item[(2-3)]Take $\xi_1\xi_2^2$ as a candidate of a head term. According to the algorithm {\bf LowCand}, $\CL=\EL\cup \LList=\{\xi_1^3,\xi_1^2\xi_2\}$. In Example~\ref{linear}, $\xi_1\xi_2^2-\frac{1}{2}t\xi_1^3$ is in $H_F$. By the algorithm {\bf renew\_low}, $\LL=\{\xi_1^3\}, \LList=\{\xi_1^3\}$, $\RR=\EL\backslash \LL=\{\xi_1^2\xi_2\}$ and $\EL$ is renewed as the empty set.

\item[(2-4)]Take $\xi_2^3$ as a candidate of a head term. Then, as {\bf Neighbor}$(\LL)=\{\xi_1^4, \xi_1^3\xi_2\}$ and $\xi_2^3 \prec \xi_1^4, \xi_2^3 \prec \xi_1^3\xi_2$, we obtain $\UU=\{\xi_1^4, \xi_1^3\xi_2\}$ and $\EL=\emptyset \cup \RR=\{\xi_1^2\xi_2\}$. Hence, $\CL=\EL\cup \LList=\{\xi_1^3,\xi_1^2\xi_2\}$. Since $\xi_2^3-\frac{1}{2t}\xi_1^2\xi_2$ is in $H_F$ by Example~\ref{linear}, then $\LL=\{\xi_1^2\xi_2\}$,  $\LList=\{\xi_1^3, \xi_1^2\xi_2\}$ and $\RR=\EL\backslash \LL=\emptyset$.

\item[(2-5)]Take $\xi_1\xi_2^3$ as a candidate of a head term. Then, as {\bf Neighbor}$(\LL)=\{\xi_1^3\xi_2, \xi_1^2\xi_2^2\}$ and $\xi_1^3\xi_2 \prec \xi_1\xi_2^3, \ \xi_1^2\xi_2^2 \prec \xi_1\xi_2^3$, we obtain $D=\{\xi_1^3\xi_2, \xi_1^2\xi_2^2\}$. Moreover, $U=\{\xi_1^4, \xi_1^3\xi_2\}$. $\UU$ is renewed as $\{\xi_1^4, \xi_1^3\xi_2\}\backslash U=\emptyset$ and $\RR$ is renewed as $U\cup \RR=\{\xi_1^4, \xi_1^3\xi_2\}$. As $D\cap \RR=\{\xi_1^2\xi_2^2\}$, $\EL=D\cup\RR=\{\xi_1^4,\xi_1^2\xi_2^2,\xi_1^3\xi_2\}$ and 
$\CL=\EL\cup\LList=\{\xi_1^3, \xi_1^2\xi_2, \xi_1^4,\xi_1^3\xi_2, \xi_1^2\xi_2^2\}$. 
Since $\xi_1\xi_2^3$ can not be a head term in $H_F$ by Example~\ref{linear}, the set $\EL$ is renewed as $\EL\cup\{\xi_1\xi_2^3\}=\{\xi_1^4,\xi_1^3\xi_2, \xi_1^2\xi_2^2,\xi_1\xi_2^3\}$.

\item[(2-6)]Take $\xi_2^4$ as a candidate of a head term. By the algorithm {\bf LowCand}, 
$\CL=\EL\cup \LList=\{\xi_1^3, \xi_1^2\xi_2, \xi_1^4,\xi_1^3\xi_2,\xi_1^2\xi_2^2,\xi_1\xi_2^3\}$. 
In Example~\ref{linear}, $\xi_2^4-\frac{1}{t}\xi_1^2\xi_2^2+\xi_1^4$  is in $H_F$. By the algorithm {\bf renew\_low}, $\LL=\{\xi_1^4,\xi_1^2\xi_2^2\},$ $\LList=\{\xi_1^3, \xi_1^2\xi_2, \xi_1^4,\xi_1^2\xi_2^2\}$, $\RR=\EL\backslash \LL=\{\xi_1^3\xi_2,\xi_1\xi_2^3\}$ and $\EL$ is renewed as the empty set.

\item[(2-7)]Take $\xi_2^5$ as a candidate of a head term. Then, as {\bf Neighbor}$(\LL)=\{\xi_1^5,$ $\xi_1^4\xi_2,\xi_1^3\xi_2^2,$ $\xi_1^2\xi_2^3\}$ and $\xi_1^4\xi_2/\xi_1, \xi_1^3\xi_2^2/\xi_1, \xi_1^2\xi_2^3/\xi_1$ $\notin \het(\SList)\cup \MList\cup \LList$, hence, by the algorithm {\bf LLem}, $D=\{\xi_1^5\}$. Moreover, $\EL=D\cup\RR=\{\xi_1^3\xi_2,\xi_1\xi_2^3,\xi_1^5\}$ and $\CL=\EL\cup\LList=\{\xi_1^3, \xi_1^2\xi_2, \xi_1^3\xi_2, \xi_1^2\xi_2^2, \xi_1\xi_2^3,\xi_1^4,\xi_1^5\}$. 
\end{enumerate}

\end{exmp}

\subsection{On danger strata}\label{3.3}
Here, we present an algorithm for computing bases of algebraic local cohomology classes $H_F$, on danger strata. 
Basically, we follow the first part of the main algorithm {\bf ALCohomology} for danger strata. However, we can not directly follow it, because the termination of the algorithms {\bf MonoSafe} and {\bf BodySafe}, is not guaranteed beforehand. 
 If $\langle GP \rangle$ is not a zero-dimensional ideal in $(\ast 1)$ of {\bf MonoSafe}, then a number of elements which do not belong to $\langle G\rangle$, may not be finite. In such a case, the algorithm does not terminate, and this means that $\langle F \rangle$ is not a zero-dimensional ideal in $K[[x]]$. 
In order to resolve this matter, the following algorithm for danger strata is introduced instead of the algorithm {\bf MonoSafe}. 
The termination of following algorithm is guaranteed by the same reason of the algorithm {\bf MonoSafe}.\\

\noindent
\SEN \vspace{1mm}
\noindent 
{\bf Algorithm 11.} {\bf (MonoDanger)}\vspace{-3mm}\\
\SEN \vspace{1mm}
\noindent 
{\bf Specification: MonoDanger(}$\A, GP${\bf )}\\
 Computing monomial elements of bases of $H_F$ on a danger stratum $\A$. \\
{\bf Input:} $(\A, GP)$: a segment of a CGS of $F$ s.t. for all $\bar{a} \in \A$, $\langle \sigma_{\bar{a}}(F)\rangle$ is nonzero-dimensional in $\bar{K}[x]$. \\
{\bf Output:} ${\cal M}$ : a finite set of triples $(\A',M,G)$ where $M, G \subset K[\xi]$. If $\langle G\rangle$ is zero-dimensional in $K[\xi]$, then the set $M$ is $\MList$ of $H_F$ on $\A'$, otherwise, $M=\emptyset$. 
 \ \\
{\bf BEGIN}\\
${\cal M}\gets \emptyset$; \ 
$B \gets$ compute a CGS of $\Mono(\CV(GP))$ on $\A$ in $K[\xi]$\\
{\bf while} $B\neq \emptyset$ {\bf do} \\
\hspace*{4mm}select $(\A',GP')$ from $B$; \ $B\gets B\backslash \{(\A',GP')\}$; \ $G\gets \het(GP')$ \\
\hspace*{4mm}{\bf if} $\dim(\langle G \rangle)= 0$ in $K[\xi]$ {\bf then}\\
\hspace*{8mm}$M\gets $ compute monomial elements which do not belong to $\langle G \rangle$ \\
\hspace*{8mm}${\cal M} \gets {\cal M} \cup \{(\A',M,G)\}$ \\
\hspace*{4mm}{\bf else} \ ${\cal M} \gets {\cal M} \cup \{(\A',\emptyset,G)\}$ \\
\hspace*{4mm}{\bf end-if}\\
{\bf end-while}\\
{\bf return} \ ${\cal M}$ \\
{\bf END}\vspace{-3mm} \\
\SEN \vspace{-3mm} \ \\

The termination of the algorithm {\bf BodySafe} is also a matter of grave concern on danger strata. We have two ideas to resolve this matter.

The first idea is preparing a natural number $\nu$ which is an estimated bound of a dimension of the vector space $H_F$. In many cases, a natural number $\nu$ can be computed from the input $F$. 
For instance, if $ f $ is a Newton non-degenerate polynomial defining an isolated singularity at the origin $ {\cal O} $, a bound of the dimension $ H_F $ can be computed by the Kouchnirenko formula \citep{Kou76}, where $ F = \{\frac{\partial f}{\partial x_1},\ldots,\frac{\partial f}{\partial x_n}\}$. 
If a number of elements of $\MList \cup \SList$ is bigger than and equal to $\nu$, then $\langle F \rangle$ is not zero-dimensional. Otherwise, $\langle F \rangle$ is zero-dimensional in $K[[x]]$.

After deciding the number $\nu$, we can compute bases of algebraic local cohomology $H_F$ on danger strata as follows.

We name the same name ({\bf BodyDanger}) to both Algorithm 12 and 13. By one's strategy, one can select one of them.\\

\noindent
\SEN \vspace{1mm}
\noindent 
{\bf Algorithm 12.} {\bf (BodyDanger)} {\sf the first idea} \vspace{-3mm}\\
\SEN \vspace{1mm}
\noindent 
{\bf Specification: BodyDanger(}$\nu, \mathcal{DL},F${\bf )}\\
Computing bases of a vector space $H_F$ on danger strata\\
{\bf Input:} $\mathcal{DL}$: a set of lists $([\A,\CT,\GList,T^{(d)},\SList,\MList,\LList, \FL,\LL,\EL,$ $\RR,\UU])$, \ $\nu$: a natural number (an estimated bound).\\ 
{\bf Output:} $(\mathcal{S},\mathcal{D})$: $\mathcal{S}=\bigcup_{i}\{[\A_i,\SList_i,\MList_i,\FL_i]\}$ where $\SList_i\cup \MList_i$ is a basis of $H_F$ on $\A_i$, and $\FL_i$ is a set of failed candidates for head terms on $\A_i$. $\mathcal{D}$ is a set of stratum. On a stratum of $\mathcal{D}$, $F$ does not satisfy $\{a\in X|f_1(a)=\cdots=f_p(a)=0\}=\{O\}$.\\
{\bf BEGIN}\\
$\mathcal{S} \gets \emptyset$; \ $\mathcal{D}\gets \emptyset$\\
{\bf while} $\mathcal{DL}\neq \emptyset$ {\bf do}\\
select $\mathcal{E}=[\A,\CT,\GList,T^{(d)},\SList,\MList,\LList, \FL,\LL,\EL, \RR,\UU]$ from $\mathcal{DL}$\\ 
$\mathcal{DL} \gets \mathcal{DL} \backslash\{\mathcal{E}\}$ \\
$N\gets$ a number of elements of $\SList\cup\MList$ \renewcommand{\arraystretch}{1}\\
\hspace*{4mm}{\bf if} $k\ge N$ {\bf then}\\
\hspace*{4mm}\begin{tabular}{|l|}\hline
($\diamondsuit$1) of {\bf BodySafe}\hspace{6.7cm} \ \ \ \ \ \ \ \ \ \ \\\hline\hline
 ($\diamondsuit$2) of {\bf BodySafe}\\\hline\hline
($\diamondsuit$3) of {\bf BodySafe}\\\hline
\end{tabular}\renewcommand{\arraystretch}{1}\\
\hspace*{4mm}$(\mathcal{S}_1,\mathcal{D}_1)\gets \text{{\bf BodyDanger}}(\nu,\mathcal{P},F)$; \ $\mathcal{S}\gets \mathcal{S}\cup\{\mathcal{S}_1\}$; \ $\mathcal{D}\gets\mathcal{D}\cup\{\mathcal{D}_1\}$\\
\hspace*{4mm}{\bf else} \ $\mathcal{D}\gets \mathcal{D}\cup \{\A\}$\\
\hspace*{4mm}{\bf end-if}\\
{\bf while-end}\\
{\bf return} $(\mathcal{S},\mathcal{D})$\\
{\bf END}\vspace{-3mm} \\
\SEN \vspace{-3mm} \ \\

The second ideal is the following. Let $\mathfrak{m}$ be the maximal ideal at the origin $O$ (i.e., $\mathfrak{m}=\langle x_1,\ldots,x_n \rangle$) and $\nu$ be a sufficient big positive integer. Then, $F_{\mathfrak{m}^{(\nu)}}=\langle F \rangle +\mathfrak{m}^{(\nu)}$ is an ideal supported at the origin $O$ where $\mathfrak{m}^{(\nu)}=\langle  x_1^\nu, x_2^\nu, \ldots,x_n^\nu \rangle$. Therefore, bases of the vector space $H_{F_{\mathfrak{m}^{(\nu)}}}$ can be computed by the algorithm {\bf BodySafe}. 
If $H_{F_{\mathfrak{m}^{(\nu)}}}=H_{F_{\mathfrak{m}^{(\nu+1)}}}$ on a stratum $\A$, it is obvious that $\langle F \rangle = F_{\mathfrak{m}^{(\nu)}}$ on $\A$. That is, if there exists $\nu \in \N$ such that $H_{F_{\mathfrak{m}^{(\nu)}}}=H_{F_{\mathfrak{m}^{(\nu+1)}}}$ on $\A$, then $\langle F \rangle$ is zero-dimensional on $\A$. If $H_{F_{\mathfrak{m}^{(\nu)}}}\neq H_{F_{\mathfrak{m}^{(\nu+1)}}}$ on $\A$, there exist some local cohomology classes in a basis of $H_{F_{\mathfrak{m}^{(\nu+1)}}}$ such that the local cohomology classes do not belong to $H_{F_{\mathfrak{m}^{(\nu)}}}$. By analyzing such local cohomology classes, we can easily guess and prove that $H_F$ has infinite many (systematic) elements which are linearly independent, on $\A$. That is, in this case, $\langle F \rangle$ is not zero-dimensional on $\A$. \\

\noindent
\SEN \vspace{1mm}
\noindent 
{\bf Algorithm 13.} {\bf (BodyDanger)} {\sf the second idea} \vspace{-3mm}\\
\SEN \vspace{1mm}
\noindent 
{\bf Specification: BodyDanger(}$\nu, \mathcal{DL},\{f_1,\ldots,f_p\}${\bf )}\\
Computing bases of a vector space $H_F$ on danger strata\\
{\bf Input:} $\mathcal{DL}$: a set of lists $([\A,\CT,\GList,T^{(d)},\SList,\MList,\LList, \FL,\LL,\EL,$ $\RR,\UU])$, \ $\nu$: a natural number (a sufficient big number).\\ 
{\bf Output:} $(\mathcal{S},\mathcal{D})$: $\mathcal{S}=\bigcup_{i}\{[\A_i,\SList_i,\MList_i,\FL_i]\}$ where $\SList_i\cup \MList_i$ is a basis of $H_F$ on $\A_i$, and $\FL_i$ is a set of failed candidates for head terms on $\A_i$. $\mathcal{D}$ is a set of stratum. On a stratum of $\mathcal{D}$, $F$ does not satisfy $\{a\in X|f_1(a)=\cdots=f_p(a)=0\}=\{O\}$.\\
{\bf BEGIN}\\
$\mathcal{D}\gets \emptyset$; $F_{\mathfrak{m}^{(\nu)}}\gets \{f_1,\ldots, f_p, x_1^\nu, x_2^\nu,\ldots, x_n^{\nu}\}$ ; $F_{\mathfrak{m}^{(\nu+1)}}\gets \{f_1,\ldots, f_p, x_1^{\nu+1}, x_2^{\nu+1},\ldots, x_n^{\nu+1}\}$ \\
${\cal H}_1 \gets${\bf SafeBody}$({\cal DL},F_{\mathfrak{m}^{(\nu)}})$ ; \ ${\cal H}_2 \gets${\bf SafeBody}$({\cal DL},F_{\mathfrak{m}^{(\nu+1)}})$ \\
${\cal S}\gets {\cal H}_1 \cap {\cal H}_2$; \ ${\cal D}_1\gets {\cal H}_1\backslash {\cal S}_1$ \\
{\bf while} ${\cal D}_1\neq \emptyset$ {\bf do} \\
\hspace*{4mm}select $\mathcal{E}=[\A',\SList',\MList',\FL']$ from $\mathcal{D}_1$; \ $\mathcal{D}_1\gets \mathcal{D}_1\backslash \{\mathcal{E}\}$ ; \ $\mathcal{D}\gets \mathcal{D}\cup\{\A'\}$\\
{\bf end-while}\\
{\bf return} $(\mathcal{S},\mathcal{D})$\\
{\bf END}\vspace{-3mm} \\
\SEN \vspace{-3mm} \ \\

We illustrate the second idea with the following example.

\begin{exmp}\normalfont 
Let us consider Example~\ref{dim}, again.  The term order is the total degree lexicographic term order such that $\xi_1\prec \xi_2$. If the parameter $t$ belongs to $\V(t-2)$ or $\V(t+2)$, the ideal $\langle F \rangle$ is not zero-dimensional in $K[x]$. Set $\nu=4$ and $F_{\mathfrak{m}^{(4)}}=\{\frac{\partial f}{\partial x_1},\frac{\partial f}{\partial x_2}, x_1^4, x_2^4\}$ and  $F_{\mathfrak{m}^{(5)}}=\{\frac{\partial f}{\partial x_1},\frac{\partial f}{\partial x_2}, x_1^5, x_2^5\}$. We apply the algorithm {\bf BodySafe} for $\{[\V(t-2),\CT, $ $\{\{\xi_1^2\xi_2,\xi_1\xi_2^2,$ $\xi_1^2\xi_2,$ $\xi_1^3\}\},T^{(d)},\SList,$ $\MList,\LList,$ $\FL,\LL,\EL,$ $\RR,\UU]\}$ with $F_{\mathfrak{m}^{(4)}}, F_{\mathfrak{m}^{(5)}}$, where $\CT, T^{(d)}, \SList, \MList,$ $\LList,$ $\FL, \LL,$ $\EL, \RR, \UU$ are empty sets. Then, a set $G_1=\{1,\xi_2,\xi_1,\xi_2^2,\xi_1\xi_2,\xi_1^2,\xi^2\xi_2-\xi_2^3, \xi_1^3-\xi_1\xi_2^2, \xi_1^3\xi_2-\xi_1\xi_2^3\}$ is a basis of the vector space $H_{F_{\mathfrak{m}^{(4)}}}$,  a set 
$G_2=\{1,\xi_2,\xi_1,\xi_2^2,\xi_1\xi_2,\xi_1^2,\xi_1^2\xi_2-\xi_2^3, \xi_1^3-\xi_1\xi_2^2, \xi_1^3\xi_2-\xi_1\xi_2^3, \xi_1^4-\xi_1^2\xi_2^2+\xi_2^4\}$ 
 is a basis of the vector space $H_{F_{\mathfrak{m}^{(5)}}}$ and $G_1\neq G_2$. 

Our implementation can compute  bases of $H_{F_{\mathfrak{m}^{(6)}}}$ and of $H_{F_{\mathfrak{m}^{(7)}}}$, too. One can find some systematic elements based on a certain rule from these bases, and guess that the following four sets \\
$\{\sum_{i=0}^{k/2}(-1)^i\xi_1^{k-2i}\xi_2^{2i}|k=2n+4, n\in \N\}$, \ $\{ \sum_{i=0}^{k/2}(-1)^i\xi_1^{k-2i}\xi_2^{2i+1}|k=2n+4, n\in \N\}$,\\
$\{\sum_{i=0}^{k/2}(-1)^i\xi_1^{k-2i+1}\xi_2^{2i}|k=2n+4, n\in \N\}$, \ $\{\sum_{i=0}^{k/2}(-1)^i\xi_1^{k-2i+1}\xi_2^{2i+1}|k=2n+4, n\in \N\}$, \\
are included in a basis of $H_F$ on $\V(t-2)$. This can be easily proved. Therefore, $\langle F \rangle$ is not zero-dimensional on $\V(t-2)$.  
One can also easily verify the non zero-dimensionality of $\langle F \rangle$ on $\V(t+2)$. 

In Fig. 6, we represent an monomial element of $H_F$ as $\bullet$ and an elements of head terms of the systematic elements as $\triangle $. 

\begin{figure}[ht]
\begin{center}
\setlength{\unitlength}{1truecm}
\begin{picture}(4.5,3.0)(0,0)
\put(0,0){\vector(1,0){4}}
\put(0,0){\line(0,1){1}}
\put(0,0){\line(1,0){4}}
\put(0,1.5){\line(1,0){0.5}}
\put(0.5,1){\line(0,1){0.5}}
\put(0.5,1){\line(1,0){3.5}}

\put(0,0){\vector(0,1){2.5}}
\put(-0.3,-0.3){$(0,0)$}
\put(0,0){$\bullet $}
\put(0,0.5){$\bullet $}
\put(0,1){$\bullet $}

\put(0.5,0){$\bullet $}
\put(0.5,0.5){$\bullet $}
\put(1,0.5){$\triangle $}
\put(1.5,0.5){$\triangle $}
\put(2,0.5){$\triangle $}
\put(2.5,0.5){$\triangle $}
\put(3,0.5){$\cdots$}
\put(3.5,0.5){$\cdots$}

\put(1.5,0.05){$\triangle $}
\put(2,0.05){$\triangle $}
\put(2.5,0.05){$\triangle $}
\put(3,0.05){$\triangle $}
\put(3.5,0.05){$\cdots$}

\put(1,0){$\bullet $}
\put(-0.35,2.1){$\xi_2$}
\put(4,-0.3){$\xi_1$}
\put(1.7,-0.9){\text{Fig. 6}}
\put(1.0,-0.5){{\it exponents}}
\end{picture}
\end{center}
\vspace{10mm}

\end{figure}
\end{exmp}

\begin{exmp}\label{avoid}\normalfont

Let $f_1=x_1^2+x_2^3+sx_2^2x_3+tx_2x_3^2, f_2=x_2^3+x_3^3$. It is described in \citep{Alek} that $ f_1=f_2=0 $ defines a quasi-homogeneous complete intersection isolated singularity provided that the parameters $ s, t $ do not belong to $ \V((s+t)^3 +(s+1)^2)) $ and the Milnor number is equal to 16. 

Let $f_3=3x_2^2x_3^2+2sx_2x_3^3+bx_3^4-sx_2^4-2tx_2^3x_3, f_4=x_1x_3^2, f_5=x_1x_2^2$ and set  $ F=\{ f_1,f_2,f_3,f_4,f_5 \}$. Since $ f_1, f_2 $ are quasi-homogeneous, a result of \citep{Greuel} on Milnor number and the Grothendieck local duality theorem \citep{Grot} imply that $ H_F $ is a vector space of dimension 16 provided $ f_1=f_2=0 $ has an isolated singularity at the origin.
 However, the algorithm {\bf ZeroDimension} outputs $\V(s-t-1)\backslash \V(t^3+2t^2+2t+1,s-t-1)$ as a danger stratum and {\bf BodyDanger} (our implementation) judges $\{a\in X|f_1(a)=f_2(a)=\cdots=f_5(a)=0\}\neq \{O\}$ on the stratum. One can check the fact $\V(s-t-1)\backslash \V(t^3+2t^2+2t+1,s-t-1) \not\subset \V((s+t)^3+(s+1)^3)$. 
For instance, take $(s,t)=(\frac{1}{2},-\frac{1}{2})\in \V(s-t-1)\backslash \V(t^3+2t^2+2t+1,s-t-1)$, then $(\frac{1}{2},-\frac{1}{2}) \notin \V((s+t)^3+(s+1)^3)$ and $\{a\in X|\sigma_{(\frac{1}{2},-\frac{1}{2})}(f_1)(a)=\cdots=\sigma_{(\frac{1}{2},-\frac{1}{2})}(f_5)(a)=0\}\neq\{O\}$. 

The algorithm {\bf BodyDanger} works powerfully to find strata on which  $\{a\in X|f_1(a)=\cdots=f_p(a)=0\}\neq\{O\}$.

\end{exmp}

We conclude this section by briefly discussing the effectiveness of the proposed method. 
In order to detect unnecessary strata on which $\langle F \rangle$ is not zero-dimensional in $K[[x]]$, first, the algorithm {\bf ZeroDimension} decomposes the parameter space $\bar{K}^m$ into safe strata and danger strata. If there exist danger strata, second, the algorithm {\bf MonoDanger} detects unnecessary strata. After that if there still exist undeterminable strata, finally, the algorithm {\bf BodyDanger} detects unnecessary strata by computing local cohomology classes of $H_F$ on the strata. The final step is a practical method for detecting unnecessary strata. Note that if we compute a parametric local cohomology system without the algorithm {\bf ZeroDimension}, then the algorithm {\bf BodyDanger} (the general case is the second idea) has to be always performed because all strata of $\bar{K}^m$ are regarded as danger. 
As we described above, {\bf BodyDanger} actually computes local cohomology classes several times. Thus, in this case, the computational complexity increases. To avoid an increase in computation cost, we have innovated the algorithm {\bf ZeroDimension}. As we described in Example~\ref{avoid}, the algorithm {\bf ZeroDimension} powerfully helps for checking unnecessary strata, and makes the computational method of a parametric local cohomology system, more effective in computational speed and complexity.

\section{Parametric standard bases}
Here, we introduce an algorithm for computing parametric standard bases of zero-dimensional ideals by using bases of algebraic local cohomology classes. 

\begin{defn}[inverse orders]\normalfont 
Let $\prec$ be a local or global term order. Then, the inverse order $\prec^{-1}$ of $\prec$ is defined by 
$x^\alpha \prec x^\beta \iff x^\beta \prec^{-1} x^\alpha$.

\end{defn}
If $\prec$ is a global term order (1 is the minimal term), then $\prec^{-1}$ is the local term order (1 is the maximal term). Conversely, if $\prec$ is a local term order, then $\prec^{-1}$ is the global term order.

\subsection{Parametric standard bases}
\begin{defn}\normalfont 
Let $F$ be a subset of $(K[t])[[x]]$, 
${\A}_i$ a stratum in ${\bar{K}}^m$, $S_i$ a subset of $K[t]_{\A_i}[[x]]$ and $\prec$ a local term order where $1\le i\le l$. 
A finite set ${\cal S}=\{({\A}_1,S_1),$ $\ldots, ({\A}_l, S_l)\}$ of 
pairs is called a {\bf parametric standard basis} on ${\A}_1\cup \cdots \cup {\A}_l$ for $\langle F\rangle$ w.r.t. $\prec$ if $\sigma_{\bar{a}}(S_i)$ is a standard basis of the ideal $\langle \sigma_{\bar{a}}(F) \rangle$ in $\bar{K}[[x]]$ w.r.t. $\prec$  
for each $i=1,\ldots,l$ and $\bar{a} \in {\A}_i$. 
\end{defn}

 Let $F=\{f_1,\ldots,f_p\}$ be a set of polynomials in $(K[t])[x]$ such that {\bf generically} $\{a\in X|$ $f_1(a)=\cdots=f_p(a)=0\}=\{O\}$ where $X$ is a neighborhood of the origin $O$ of $K^n$. Then, by utilizing the information of bases of $H_F$, one can obtain parametric standard bases of $\langle F \rangle$ in $K(t)[[x]]$.

Let us recall that there is a natural pairing, denote by $\text{res}_{\{O\}}(\ ,\ )$, between the quotient space $K[[x]]/\langle P \rangle$ and the vector space $H_{P}$ where $\langle P\rangle \subset K[[x]]$ is a zero-dimensional ideal.
$$\text{res}_{\{O\}}(\ ,\ ): K[[x]]/\langle P \rangle\times H_{P}\longrightarrow K.$$
Since the pairing is non-degenerate according to the Grothendieck local duality theorem \citep{Grot1}, we have the following result.

\begin{lem}[\cite{TN09}]\label{res}\normalfont 
Let $P=\{g_1,\ldots,g_q\}$ be a set of polynomials in $K[x]$ such that $\{a\in X|$ $g_1(a)=\cdots=g_q(a)=0\}=\{O\}$. Then, a given formal power series $h \in K[[x]]$ is in the ideal $\langle P \rangle $  if and only if for all $\varphi \in \Psi$, $h$ satisfies $\text{res}_{\{O\}}(h ,\varphi)=0$ where $\Psi$ is a basis of the vector space $H_{P}$.
\end{lem}

One can extend this fact into the parametric cases. 
The next theorem gives us the relation between bases of $H_F$ and parametric standard bases of $\langle F \rangle$. 

\begin{notation}\label{tra}\normalfont 
Let $\SList$ be a set of polynomials in $(K(t))[\xi]$ and $\LList$ be a set of lower terms of all elements of $\SList$. Suppose that there is no monomial in $\SList$, and $\SList$ has an element whose form is $\xi^{\lambda}+\sum_{\kappa  \prec  \lambda}c_{(\lambda, \kappa)}\xi^{\kappa}$ where $c_{(\lambda, \kappa)} \in K(t)$. Then, the transfer $\text{{\it SB}}_{(\SList,\LList)}$ is defined by the following:
$$\left\{
\begin{array}{lll}
\displaystyle \text{{\it SB}}_{(\SList,\LList)}(\xi^{\lambda})= x^{\lambda}-\sum_{\xi^\kappa \in \het(\SList)}c_{(\kappa,\lambda)}x^{\kappa} &\text{ in } K(t)[x] & \text{ if } \xi^{\lambda}\in \LList, \\
\text{{\it SB}}_{(\SList,\LList)}(\xi^{\lambda})=x^{\lambda} & \text{ in } K(t)[x]&  \text{ if } \xi^{\lambda}\notin \LList.
\end{array}
\right.
$$
Let $G$ be a set of terms in $K[\xi]$. Then, the set $\text{{\it SB}}_{(\SList,\LList)}(G)$ is also defined by 
$\text{{\it SB}}_{(\SList,\LList)}(G)=\{\text{{\it SB}}_{(\SList,\LList)}(\xi^{\lambda})|\xi^{\lambda}\in G\}.$
\end{notation}
\begin{thm}\label{sb}\normalfont 
Let $\prec$ be a global total degree lexicographic term order (Definition~\ref{monoorder}). 
 Let $(\mathcal{S},\mathcal{D})$ be an output of {\bf ALCohomology}$(F)$ and a list $[\A,\SList,$ $\MList,\LList, \FL]$ is in $\mathcal{S}$.  Then, for all $\bar{a}\in \A$, $\sigma_{\bar{a}}(\text{{\it SB}}_{(\SList,\LList)}(\FL))$ is the reduced standard basis of $\langle \sigma_{\bar{a}}(F) \rangle$ w.r.t. $\prec^{-1}$ (the local total degree lexicographic term order), in $\bar{K}[[x]]$. Namely, $\{(\A,\text{{\it SB}}_{(\SList,\LList)}(\FL))\}$ is a parametric standard basis on $\A$ for $F$. (The notation $\sigma$ is from section 3.1.)
\end{thm}
\begin{pf}
Since the algorithm {\bf BodySafe} decides linear combination elements of a basis of $H_F$ from bottom to up w.r.t. $\prec$ and $\langle F\rangle$ is zero-dimensional on $\A$, the set $\CV^{-1}(\FL)$ (failed candidates of head terms) becomes a set of head terms of the standard basis w.r.t. $\prec^{-1}$, on $\A$. By Lemma~\ref{res} (and Theorem 7, Proposition 8 and Theorem 9 of the paper \citep{TN09}), for all $\bar{a} \in \A$, it is obvious that if $\xi^\lambda \in \FL$ is not in $\LList$, then the monomial $x^\lambda$ itself is in the ideal $\langle \sigma_{\bar{a}}(F) \rangle$ in $\bar{K}[[x]]$, and if $\xi^\lambda \in \FL$ is in $\LList$, then $\displaystyle \sigma_{\bar{a}}\left(x^{\lambda}-\sum_{\xi^\kappa \in \het(\SList)}c_{(\kappa,\lambda)}x^{\kappa}\right)$ is also in $\langle \sigma_{\bar{a}}(F) \rangle$ and $\xi^\lambda$ is not in $\het(\SList)$ w.r.t. $\prec$. Hence, for all $\bar{a} \in \A$, $\sigma_{\bar{a}}\left(\text{{\it SB}}_{(\SList,\LList)}(\FL)\right)$ is the reduced standard basis of $\langle \sigma_{\bar{a}}(F)\rangle$ w.r.t. $\prec^{-1}$ on $\A$.
\end{pf}

This theorem leads us to construct the following algorithm for computing parametric standard bases.\\

\noindent
\SEN \vspace{1mm}
\noindent 
{\bf Algorithm 14.} {\bf (StandardBases1)}\vspace{-3mm}\\
\SEN \vspace{1mm}
\noindent 
{\bf Specification: StandardBases1(}$F${\bf )}\\
Computing a parametric standard basis for a zero-dimensional ideal  $\langle F\rangle$. \\
{\bf Input:} $F \in (K[t])[x]$, $\prec$: a global total degree lexicographic term order. \\
{\bf Output:} $(\mathcal{S},\mathcal{A}_2)$: $\mathcal{S}$ is a set of pairs $(\A,E)$ such that for all $\bar{a}\in \A$, $\sigma_{\bar{a}}(E)$ is the reduced standard basis of $\langle \sigma_{\bar{a}}(F)\rangle$ w.r.t. $\prec^{-1}$. $\mathcal{A}_2$ is described in the algorithm  {\bf ALCohomology}.\\
{\bf BEGIN}\\
$\mathcal{S}\gets \emptyset$; \ $(\mathcal{A}_1,\mathcal{A}_2)\gets$ {\bf ALCohomology}$(F)$ \\
{\bf while} $\mathcal{A}_1 \neq \emptyset$ {\bf do} \\
\hspace*{4mm}select $\mathcal{B}=[\A',\SList,\MList,\LList,\FL]$ from $\mathcal{A}_1$; \ $\mathcal{A}_1\gets \mathcal{A}_1\backslash \{\mathcal{B}\}$\\
\hspace*{4mm}$\mathcal{S}\gets\mathcal{S}\cup \{(\A',SB_{(\SList,\LList)}(\FL))\}$\\
{\bf end-while}\\
 {\bf return}$(\mathcal{S},\mathcal{A}_2)$\\
{\bf END}\vspace{-3mm} \\
\SEN \vspace{-3mm} \ \\

This algorithm gives a nice decomposition of the parameter space depending on structure of bases of $H_F$ thanks to the algorithm {\bf ALCohomology}. This is the big advantage of this algorithm.

\begin{exmp}\normalfont 
Let $f=x_1^4+tx_1^2x_2^2+x_2^4$ be a polynomial with a parameter $t$ in $(\C[t])[x_1,x_2]$ and $\prec$ be the global total degree lexicographic term order such that $\xi_1 \prec \xi_2$. Set $F=\{\frac{\partial f}{\partial x_1}, \frac{\partial f}{\partial x_2}\}$. The output of {\bf ALCohomology}$(F)$, is already given in Example~\ref{linear}.
\begin{enumerate} 
\item[(i)]On $\V(t)$, $\SList=\emptyset, \MList=\{1,\xi_1,\xi_2,\xi_1^2,\xi_1\xi_2,\xi_2^2,\xi_1^2\xi_2,\xi_1\xi_2^2,\xi_1^2\xi_2^2\}$, $\LList=\emptyset$ and $\FL=\{\xi_1^3, \xi_2^3\}$. 
\item[(ii)]On $\C^2\backslash \V(t(t^2-4))$, $\SList=\{\xi_1\xi_2^2-\frac{1}{2}t\xi_1^3, \xi_2^3-\frac{1}{2t}\xi_1^2\xi_2, \xi_2^4-\frac{1}{t}\xi_1^2\xi_2^2+\xi_1^4\}$, $\MList=\{1,\xi_1,\xi_2,\xi_1^2,\xi_1\xi_2,$ $\xi_2^2\}$, $\LList=\{\xi_1^3, \xi_1^2\xi_2, \xi_1^2\xi_2^2, \xi_1^4\}$ and $\FL=\{\xi_1^3, \xi_1^2\xi_2,\xi_1\xi_2^3,\xi_2^5\}.$
\end{enumerate}
In case (i), $\text{{\it SB}}_{(\SList,\LList)}(\FL)=\{x_1^3, x_2^3\}$ is the reduced standard basis of $\langle F \rangle$ w.r.t. $\prec^{-1}$. \\
In case (ii), each elements of $\FL$ is transformed as follows: 
\begin{center}
$\xi_1^3$ \ $\longrightarrow$ \  $x_1^3+\frac{t}{2}x_1x_2^2$,  \ \ 
$\xi_1^2\xi_2$ \ $\longrightarrow$ \  $x_1^2x_2+\frac{2}{t}x_2^3$, \ \ 
$\xi_1\xi_2^3$ \  $\longrightarrow$ \  $x_1x_2^3$, \ \
$\xi_2^5$ \ $\longrightarrow$ \  $x_2^5$.
\end{center}

Therefore, $\{x_1^3+\frac{t}{2}x_1x_2^2, x_1^2x_2+\frac{2}{t}x_2^3, x_1^3x_2, x_1^5\}$ is the reduced standard basis of $\langle F \rangle$ w.r.t $\prec^{-1}$  on $\C\backslash \V(t(t^2-4))$. 

Let us remark that if $t=\pm 2$, then $\langle F \rangle$ is not zero-dimensional in $K[[x]]$.
\end{exmp}

All algorithms of this paper have been implemented in the computer algebra system \verb|Risa/Asir| by the authors. In the following example, we give an output of our implementation.
\begin{exmp}\normalfont 
Let $F=\{3sx_1^2+2x_1x_2^2+tx_2^3,2x_1^2x_2+5x_2^4+3tx_1x_2^2\}$ be a set of polynomials with parameters $s, t$ in $(\C[s,t])[x_1, x_2]$, and $\prec$ be the global total degree lexicographic term order such that $x_1 \prec x_2$. Generically, $F$ has only the point $O$ in $X$ where $X$ is a neighborhood of the origin $O$ of $\C^2$. The variables $\xi_1, \xi_2$ are corresponding to variables $x_1, x_2$. Our implementation outputs bases of the vector space $H_F$ and standard bases of $\langle F\rangle$ on $\C^2$ w.r.t $\prec^{-1}$, as follows. 
\begin{enumerate}
\setlength{\leftskip}{-0.3cm}
\item[-]If the parameters belong to $\V(s)$, then $\langle F \rangle$ is not zero-dimensional. 
\item[-] If the parameters belong to $\C^2\backslash \V(st(-15s+2t))$, then a set $\left\{1, \xi_2, \xi_1,\xi_2^2,\xi_1\xi_2, s\xi_2^3-\right.$ $\left. \frac{1}{3}t\xi_1^2, s^2t\xi_2^4+(-\frac{5}{3}s^2+\frac{2}{9}st)\xi_1\xi_2^2+(\frac{10}{9}s-\frac{4}{27}t) \xi_1^2-\frac{1}{3}st^2\xi_1^2\xi_2\right\}$ 
 is a basis of $H_F$. Hence, $\{s^2tx_1^2+(-\frac{10}{9}s+\frac{4}{27}t)x_2^4+\frac{1}{3}st^2x_2^3,$ $s^2tx_1x_2^2+(\frac{5}{3}s^2-\frac{2}{9}st)x_2^4,x_2^5\}$ is a parametric standard basis w.r.t $\prec^{-1}$.
\item[-] If the parameters belong to $\V(-15s+2t)\backslash \V(s,t)$, then a set $\left\{1, \xi_2, \xi_1, \xi_2^2, \xi_1\xi_2, s\xi_2^3-\right.$ $\left.\frac{1}{3}t\xi_1^2, s\xi_2^4-\frac{1}{3}t\xi_1^2\xi_2\right\}$
 is a basis of $H_F$. Hence, $\{sx_1^2+\frac{1}{3}tx_2^3,x_1x_2^2,x_2^5\}$ 
 is a parametric standard basis w.r.t $\prec^{-1}$. 
\item[-]If the parameters belong to $\V(t)\backslash \V(s,t)$, then a set $\left\{1, \xi_2, \xi_1,\xi_2^2,\xi_1\xi_2, \xi_2^3, s\xi_1\xi_2^2-\frac{2}{3}\xi_1^2,\right.$ $\left. \frac{4}{15}\xi_2^4+s\xi_1\xi_2^3-\frac{2}{3}\xi_1^2\xi_2\right\}$ 
 is a basis of $H_F$. Hence, $\{sx_1^2+\frac{2}{3}x_1x_2^2,-\frac{4}{15}x_1x_2^3+sx_2^4\}$ is a parametric standard basis w.r.t $\prec^{-1}$. 
\end{enumerate}
Note that in case $\V(t)\backslash \V(s,t)$, the dimension of the vector space $H_F$ is 8, but in cases $\C^2\backslash \V(st(-15s+2t))$ and $\V(-15s+2t)\backslash \V(s,t)$, the dimension of the vector spaces $H_F$ are 7. Our implementation tells us this deference.
\end{exmp}


\subsection{Other local term orders}
The algorithm {\bf ALCohomology} has been constructed based on a global total  degree lexicographic term order $\prec_1$. That's why reduced standard bases w.r.t. $\prec_1^{-1}$, are directly obtained by outputs of the algorithm {\bf ALCohomology}. Here, we describe how to compute standard base w.r.t. other local term orders.

Let $({\cal S}, {\cal D})$ be an output of {\bf ALCohomology}$(F)$ and $\prec$ be a global term order in $K[\xi]$. Suppose that $[\A, \SList, \MList, \LList, \FL]\in {\cal S}$, $\SList=\{\psi_1,\ldots, \psi_\rho\}$ $\subset K[\xi]$ and the list $[\xi^{\alpha_1},\xi^{\alpha_2},$ $\ldots, \xi^{\alpha_r}]$ is lined up all elements of $\Mono(\SList)$ in order of $\prec$ where $\xi^{\alpha_r} \prec \xi^{\alpha_{n-1}} \prec \cdots\prec  \xi^{\alpha_1}$. Moreover, let $M$ the coefficient matrix of $\SList$ w.r.t. the vector $\ ^t(\xi^{\alpha_1}, \xi^{\alpha_2}, \ldots, \xi^{\alpha_r})$ (i.e., $\ ^t(\psi_1, \ldots, \psi_\rho)=M\ ^t(\xi^{\alpha_1}, \xi^{\alpha_2},$ $\ldots, \xi^{\alpha_r})$) where $\ ^t(\psi_1, \ldots, \psi_\rho)$ is the transposed matrix of $(\psi_1, \ldots, \psi_\rho)$. Then, it is possible to compute the row reduced echelon matrix of $M$ on $\A$, like a method for solving the system of parametric linear equations. Let $\{(\A_1,M_1'),(\A_2,M_2'),\ldots, (\A_l,M_l')\}$ be a set of pairs such that for each $1\le i \le l$, $M_i'$ is the row reduced echelon matrix of $M$ on $\A_i$ and $\A=\A_1\cup\cdots \cup \A_l$. Then, we have the following theorem.

\begin{thm}\normalfont 
Using the same notation as in above discussion, let $\ ^t(\varphi_1, \varphi_2,\cdots$ $, \varphi_\rho)=M_i'\ ^t(\xi^{\alpha_1}, \xi^{\alpha_2},$ $\ldots, \xi^{\alpha_r})$ where $1\le i \le l$. Suppose that $SL=\{\varphi_1, \varphi_2, \cdots,$ $ \varphi_\rho\}$, $L=\Mono(SL)\backslash \het(SL)$, $T=\het(SL)\cup \MList$ and the reduced Gr\"obner basis of $\langle \text{{\bf Neighbor}}(T)$ $\backslash T \rangle$ is $\FL_{\prec}$ w.r.t. $\prec$. Then, for all $\bar{a}\in \A_i$, $\sigma_{\bar{a}}(SB_{(SL,L)}(\FL_{\prec}))$ is the reduced standard basis of $\langle \sigma_{\bar{a}}(F)\rangle$ w.r.t. $\prec^{-1}$ in $\bar{K}[[x]]$. (The transfer $SB_{(SL,L)}$ is from Notation~\ref{tra}.)
\end{thm} 
\begin{pf}
As $\SList \cup \MList$ is a basis of $H_F$ on $\A$, it is obvious that $SL\cup \MList$ is a basis of $H_F$ on $\A_i$, too. $L$ is a set of lower terms of $SL$ w.r.t $\prec$. Since $M_i'$ is the row reduced echelon matrix of $M$ w.r.t. the vector $\ ^t(\xi^{\alpha_1}, \xi^{\alpha_2}, \ldots, \xi^{\alpha_r})$ on $\A$, the set $\CV^{-1}(\FL_{\prec})$ becomes a set of head terms of the standard basis w.r.t. $\prec^{-1}$ on $\A_i$. By this observation and Theorem~\ref{sb}, this theorem holds.
\end{pf}

This theorem leads us to construct the following algorithm for computing parametric standard bases w.r.t. any local term order.\\

\noindent
\SEN \vspace{1mm}
\noindent 
{\bf Algorithm 15.} {\bf (StandardBases2)}\vspace{-3mm}\\
\SEN \vspace{1mm}
\noindent 
{\bf Specification: StandardBases2(}$F, \prec${\bf )}\\
Computing a parametric standard basis for $\langle F\rangle$ w.r.t. $\prec$. \\
{\bf Input:} $F \subset (K[t])[x]$, $\prec$: a local term order, \\
{\bf Output:} $(\mathcal{S},\mathcal{A}_2)$: $\mathcal{S}$ is a set of pairs $(\A,E)$ such that for all $\bar{a}\in \A$, $\sigma_{\bar{a}}(E)$ is the reduced standard basis of $\langle \sigma_{\bar{a}}(F)\rangle$ w.r.t. $\prec$. $\mathcal{A}_2$ is described in the algorithm {\bf ALCohomology}.\\
{\bf BEGIN}\\
$\mathcal{S}\gets \emptyset$; \ $(\mathcal{A}_1,\mathcal{A}_2)\gets$ {\bf ALCohomology}$(F)$ \\
{\bf while} $\mathcal{A}_1 \neq \emptyset$ {\bf do} \\
\hspace*{4mm} select $\mathcal{B}=[\A',\SList,\MList,\LList,\text{FA}]$ from $\mathcal{A}_1$; \ $\mathcal{A}_1\gets \mathcal{A}_1\backslash \{\mathcal{B}\}$ \\
\hspace*{4mm} $v\gets$ Line up all elements of $\Mono(\SList)$ w.r.t. $\prec^{-1}$. \\
\hspace*{4mm} $M \gets$ Make the coefficient matrix of $\SList$ w.r.t. $v$ \\
\hspace*{4mm} ${\cal AM} \gets$ Compute the row reduced echelon matrix of $M$ on $\A'$\\
\hspace*{8mm} {\bf while} ${\cal AM} \neq \emptyset$ {\bf do}\\
\hspace*{8mm} select $(\A'',M')$ from ${\cal AM}$; \ ${\cal AM}\gets {\cal AM}\backslash \{(\A'',M')\}$ \\
\hspace*{8mm} (where $M'$ is the row reduced echelon matrix of $M$ on $\A''$.) \\
\hspace*{8mm} $^t(\varphi_1, \varphi_2, \cdots, \varphi_\rho)\gets M'\ ^t v$; \ $SL\gets \{\varphi_1, \varphi_2, \cdots, \varphi_\rho\}$ \\
\hspace*{8mm} $L\gets \Mono(SL)\backslash \het(SL)$; \  $T\gets \het(SL)\cup \MList$ \\
\hspace*{8mm} $\FL_{\prec}\gets$ the reduced Gr\"obner basis of $\langle \text{{\bf Neighbor}}(T)\backslash T \rangle$ \\
\hspace*{8mm} $\mathcal{S}\gets\mathcal{S}\cup \{(\A'',SB_{(SL,L)}(\FL_{\prec}))\}$\\
\hspace*{8mm} {\bf end-while}\\
\hspace*{4mm}{\bf end-while}\\
 {\bf return}$(\mathcal{S},\mathcal{A}_2)$\\
{\bf END}\vspace{-3mm} \\
\SEN \vspace{-3mm} \ \\

\begin{exmp}\normalfont 
Let $f=x_1^4+tx_1^2x_2^2+x_2^4$ be a polynomial with a parameter $t$ in $(\C[t])[x_1,x_2]$. Set $F=\{\frac{\partial f}{\partial x_1}, \frac{\partial f}{\partial x_2}\}$. The output of {\bf ALCohomology}$(F)$, is already given in Example~\ref{linear}. If a parameter $t$ belongs to $\C\backslash \V(t(t^2-4))$, then $\{1,\xi_1,\xi_2,\xi_1^2,\xi_1\xi_2,$ $\xi_2^2, \xi_1\xi_2^2-\frac{1}{2}t\xi_1^3, \xi_2^3-\frac{1}{2t}\xi_1^2\xi_2, \xi_2^4-\frac{1}{t}\xi_1^2\xi_2^2+\xi_1^4\}$ is a basis of $H_F$. 
Let $\prec$ be the local lexicographic term order such that $x_1\prec x_2$ and $SL=\{\xi_1\xi_2^2-\frac{1}{2}t\xi_1^3, \xi_2^3-\frac{1}{2t}\xi_1^2\xi_2, \xi_2^4-\frac{1}{t}\xi_1^2\xi_2^2+\xi_1^4\}$. 

We compute a parametric standard basis of $\langle F \rangle$ w.r.t. $\prec$ on $\C\backslash \V(t(t^2-4))$. First, we line up all elements of $\Mono(\SList)$ w.r.t. $\prec^{-1}$ (the global lexicographic term order), then we get the vector $v=\ ^t(\xi_1^4, \xi_1^3, \xi_1^2\xi_2^2, \xi_1^2\xi_2, \xi_1\xi_2^2, \xi_2^4, \xi_2^3)$. The coefficient matrix of $\SList$ w.r.t. $v$ is $M$, and the row reduced echelon matrix of $M$ is $M'$: 
$$
 M = \hspace{-2mm} \bordermatrix{
 & \xi_1^4 & \xi_1^3 & \xi_1^2\xi_2^2 & \xi_1^2\xi_2 & \xi_1\xi_2^2 & \xi_2^4 & \xi_2^3 \cr
 & 0 & -1/2t & 0 & 0 & 1 & 0 & 0 \cr
 & 0 & 0 & 0 & -1/2t & 0 & 0& 1 \cr
 & 1 & 0 & -1/t & 0 & 0 & 1 & 0 \cr
 }, \ 
 M' =\left( 
\begin{array}{ccccccc}
 1 & 0 & -1/t & 0 & 0 & 1 & 0 \cr
 0 & 1 & 0 & 0 & -2t & 0& 0 \cr
 0 & 0 & 0 & 1 & 0 & 0 & -2t \cr
\end{array} 
 \right)
.$$
Then, $M'\ ^tv=\ ^t(\xi_1^4-\frac{1}{t}\xi_2^2\xi_2^2, \xi_1^3-2t\xi_1\xi_2^2, \xi_1^2\xi_2-2t\xi_2^3)$. Hence, $SL=\{\xi_1^4-\frac{1}{t}\xi_2^2\xi_2^2, \xi_1^3-2t\xi_1\xi_2^2, \xi_1^2\xi_2-2t\xi_2^3\}$, $\het(SL)=\{\xi_1^4, \xi_1^3, \xi_1^2\xi_2\}$, $L=\Mono(SL) \backslash \het(SL)=\{\xi_1^2\xi_2^2,$ $\xi_1\xi_2^2, \xi_2^4, \xi_2^3\}$ and $T=\het(SL)\cup \MList$. Since the set $\FL_{\prec}=\{\xi_1^5, \xi_1^3\xi_2, \xi_1\xi_2^2, \xi_2^3\}$ is the reduced Gr\"obner basis of $\langle \text{{\bf Neighbor}}(T)\backslash T \rangle$, the parametric standard basis of $\langle F \rangle$ w.r.t. $\prec$ on $\C\backslash \V(t(t^2-4))$ is 
$$SB_{(SL,L)}(\FL_{\prec})=\{x_1^5,x_1^3x_2,x_1x_2^2+2tx_1^3, x_2^3+2tx_1^2x_2\}.$$
In Fig. 7, we represent an element of $\MList$ as $\bullet$, an element of $\het(SL)$ as $\triangle$. As the set $\FL_{\prec}$ plays a key role to construct standard bases, we give the elements of $\FL_{\prec}$ in the figure. \vspace{-3mm}

\begin{figure}[ht]
\begin{center}
\setlength\unitlength{1truecm}
\begin{picture}(4.5,2.5)(0,0)
\put(0,0){\vector(1,0){4}}
\put(0,0){\line(0,1){1}}
\put(0,0){\line(1,0){4}}
\put(0,1.5){\line(1,0){0.5}}
\put(2.5,0){\line(0,1){0.5}}
\put(1.5,0.5){\line(1,0){1}}
\put(1.5,0.5){\line(0,1){0.5}}
\put(0.5,1.0){\line(1,0){1}}

\put(0.5,1){\line(0,1){0.5}}
\put(0,0){\vector(0,1){2.5}}
\put(-0.3,-0.3){$(0,0)$}
\put(0,0){$\bullet $}
\put(0,0.5){$\bullet $}
\put(0,1){$\bullet $}

\put(0.5,0){$\bullet $}
\put(0.5,0.5){$\bullet $}
\put(1.5,0.05){$\triangle $}
\put(2.0,0.05){$\triangle $}
\put(1,0.5){$\triangle $}

\put(1,0){$\bullet $}
\put(2.55,0.13){$\xi_1^5$}
\put(0.05,1.653){$\xi_2^3$}
\put(-0.35,2.2){$\xi_2$}
\put(0.55,1.13){$\xi_1\xi_2^2$}
\put(1.55,0.63){$\xi_1^3\xi_2$}
\put(4,-0.3){$\xi_1$}
\put(1.7,-0.9){\text{Fig. 7}}
\put(1.0,-0.5){\text{exponents}}

\end{picture}
\end{center}
\end{figure}
\end{exmp}

\section{Conclusions}
We have described algorithms for computing parametric local cohomology systems, and given a new algorithm for computing parametric standard bases as an application. The algorithm for computing parametric standard bases, has the following advantages.
\begin{enumerate}
\item[-]The algorithm always outputs a ``reduced" standard basis. The computer algebra system \verb|Singular| has a command that outputs a (non-parametric) standard basis. \verb|Singular| does not have this property.

\item[-] The substantial computation consists of only linear algebra computation.
\item[-] We do not need Mora's reduction (tangent cone algorithm\citep{Mora82}) for computing standard bases.
\item[-] The algorithm outputs a nice decomposition of the parameter space depending on the structure of standard bases w.r.t. a local total degree lexicographic term order.
\end{enumerate} 

All algorithms of this paper, have been implemented in the computer algebra system \verb|Risa/Asir|. Actually, there does not exist any implementation for computing ``parametric" standard bases, except for our implementation. Only our implementation exists for them. Our implementation is useful for studying and analyzing singularities.

\bibliographystyle{elsart-harv}

\bibliography{knabe2014}

\begin{thebibliography}{22}
\expandafter\ifx\csname natexlab\endcsname\relax\def\natexlab#1{#1}\fi
\expandafter\ifx\csname url\endcsname\relax
  \def\url#1{\texttt{#1}}\fi
\expandafter\ifx\csname urlprefix\endcsname\relax\def\urlprefix{URL }\fi

\bibitem[{Aleksandrov(1983)}]{Alek}
Aleksandrov, A.~G., 1983. Normal forms of one-dimensional quasihomogeneous
  complete intersections. Mathematics of the USSR-Sbornik 45/1, 1--30.

\bibitem[{Brodmann{, }{M. P}. and Sharp{, }{R. Y}.(1998)}]{BS98}
Brodmann{, }{M. P}., Sharp{, }{R. Y}., 1998. Local {C}ohomology. Cambridge
  Univ. Press.

\bibitem[{Decker{, }W. et~al.(2012)Decker{, }W., Greuel{, }G.-M., Pfister{,
  }G., and H.}]{DGPS}
Decker{, }W., Greuel{, }G.-M., Pfister{, }G., H., S., 2012. {\sc Singular}
  {3-1-6} --- {A} computer algebra system for polynomial computations.
  \url{http://www.singular.uni-kl.de}.

\bibitem[{Gao and Chou(1992)}]{G1}
Gao, X., Chou, S., 1992. Solving parametric algebraic systems. In: Wang, P.
  (Ed.), International Symposium on Symbolic and Algebraic Computation
  (ISSAC1992). ACM-Press, pp. 335--341.

\bibitem[{Greuel{, }G.-M.(1975)}]{Greuel}
Greuel{, }G.-M., 1975. Der {G}au\ss-{M}anin-{Z}usammenhang isolierter
  singularit\"aten von vollst\"andigen durchschnitten. Math Ann. 214, 235--266.

\bibitem[{Grothendieck(1957)}]{Grot1}
Grothendieck, A., 1957. Th\'eor\`emes de dualit\'e pour les faisceaux
  alg\'ebriques coh\'erents. S\'eminaire Bourbaki 149.

\bibitem[{Grothendieck(1967)}]{Grot}
Grothendieck, A., 1967. {L}ocal {C}ohomology, notes by {R}. {H}artshorne,
  Lecture Notes in Math., \text{41}. Springer.

\bibitem[{Iarrobino(1984)}]{Iar84}
Iarrobino, A., 1984. Compressed algebras: Artin algebras having given socle
  degrees and maximal length. Trans. AMS 285, 337--378.

\bibitem[{Iarrobino and Emsalem(1978)}]{Iar78}
Iarrobino, A., Emsalem, J., 1978. Some zero-dimensional generic singularities:
  finite algebras having small tangent spaces. Compositio Math. 36, 145--188.

\bibitem[{Kapur et~al.(2010)Kapur, Sun, and Wang}]{ksd10}
Kapur, D., Sun, Y., Wang, D., 2010. A new algorithm for computing comprehensive
  {G}r\"obner systems. In: Watt, S. (Ed.), International Symposium on Symbolic
  and Algebraic Computation (ISSAC). ACM-Press, pp. 29--36.

\bibitem[{Kouchnirenko(1976)}]{Kou76}
Kouchnirenko, A., 1976. Poly\`edre de {N}ewton et nombres de {M}ilnor. Invent.
  Math. 32, 1--31.

\bibitem[{Lyubeznik{, }G.(2002)}]{Lyu02}
Lyubeznik{, }G., 2002. Local Cohomology and its Applications. Dekker.

\bibitem[{Montes(2002)}]{Mo02}
Montes, A., 2002. A new algorithm for discussing {G}r\"obner basis with
  parameters. Journal of Symbolic Computation 33/1-2, 183--208.

\bibitem[{Mora(1982)}]{Mora82}
Mora, T., 1982. An algorithm to compute the equations of tangent cones. In:
  Lecture Notes in Computer Science Vol.144. Springer, pp. 158--165.

\bibitem[{Nabeshima(2012)}]{Na12}
Nabeshima, K., 2012. Stability conditions of monomial bases and comprehensive
  {G}r\"obner systems,. In: Gerdt, V., Koepf, W., Mayr, E., Vorozhtsov, E.
  (Eds.), Computer Algebra in Scientific Computing (CASC), Lecture Notes in
  Computer Science. Vol. 7442. Springer, pp. 248--259.

\bibitem[{Nabeshima and Tajima(2014)}]{NT14}
Nabeshima, K., Tajima, S., 2014. On efficient algorithms for computing
  parametric local cohomology classes associated with semi-quasihomogeneous
  singularities and standard bases. In: Nabeshima, K. (Ed.), International
  Symposium on Symbolic and Algebraic Computation (ISSAC2014). ACM-Press, pp.
  351--358.

\bibitem[{Noro and Takeshima(1992)}]{NT92}
Noro, M., Takeshima, T., 1992. Risa/{A}sir- {A} computer algebra system. In:
  Wang{, }P. (Ed.), International Symposium on Symbolic and Algebraic
  Computation (ISSAC1992). ACM-Press, pp. 387--396.

\bibitem[{Sit(1992)}]{sit92}
Sit, W., 1992. An algorithm for solving parametric linear systems. Journal of
  Symbolic Computation 13, 353--394.

\bibitem[{Suzuki{, }A. and Sato{, }Y.(2003)}]{ss03}
Suzuki{, }A., Sato{, }Y., 2003. An alternative approach to comprehensive
  {G}r\"obner bases. Journal of Symbolic Computation 36/3-4, 649--667.

\bibitem[{Tajima and Nakamura(2009)}]{TN09}
Tajima, S., Nakamura, Y., 2009. Annihilating ideals for an algebraic local
  cohomology class. Journal of Symbolic Computation 44, 435--448.

\bibitem[{Tajima et~al.(2009)Tajima, Nakamura, and Nabeshima}]{TNN09}
Tajima, S., Nakamura, Y., Nabeshima, K., 2009. Standard bases and algebraic
  local cohomology for zero dimensional ideals. Advanced Studies in Pure
  Mathematics 56, 341--361.

\bibitem[{Weispfenning{, }V.(1992)}]{wv92}
Weispfenning{, }V., 1992. Comprehensive {G}r\"obner bases. Journal of Symbolic
  Computation 14/1, 1--29.

\end{thebibliography}

\end{document}